# Characterization and Monitoring of Nonlinear Dynamics and Chaos in Complex Physiological Systems

**Hui Yang[*1], Yun Chen[1], and Fabio Leonelli[2]**

[1]Complex Systems Monitoring, Modeling and Control Laboratory, The Pennsylvania State University, University Park, PA, USA
[2]Cardiac Electrophysiology Laboratory, James A. Haley Veterans' Hospital, Tampa, FL, USA

## ABSTRACT

Nonlinear dynamics arise whenever multifarious entities of a system cooperate, compete, or interfere. For example, cardiovascular system involves a great level of complexity. Multi-lead electrocardiogram (ECG) signals are generated through orchestrated depolarization and repolarization of cells and manifest significant nonlinear dynamics. Nonlinear dynamical systems defy understanding based on the traditional reductionist's approach, in which one attempts to understand a system's behavior by combining all constituent parts that have been analyzed separately. In order to cope with system complexity, modern healthcare systems are investing in advanced physiological sensing and patient monitoring, thereby giving rise to big data. Realizing the full potential of big data for healthcare intelligence requires fundamentally new methodologies to harness and exploit complexity. However, available nonlinear dynamics techniques are either not concerned with healthcare analytical objectives or fail to effectively analyze big data to extract useful information for improving healthcare services. There is an urgent need to develop analytical methodologies that fully exploit the underlying nonlinear dynamics in physiological systems for advancing healthcare services with exceptional features such as personalization, responsiveness, and superior quality. This chapter presents some theoretical developments and tools to advance the applications of nonlinear dynamics principles in health care. Specifically, we focus on sensor-based characterization and modeling of nonlinear dynamics (i.e., multifractal analysis and multiscale recurrence quantification). Then, current developments and applications of these methodologies are examined for characterizing and exploiting heart rate variability and space-time ECG signals. It is our expectation that this chapter will spur further development of nonlinear dynamics methodologies for improving healthcare services and accelerating the discovery of scientific knowledge in biomedical research.

### Keywords

Nonlinear dynamics, medical monitoring, recurrences, fractals, physiological systems.

[*] Corresponding author: Dr. Hui Yang, e-mail: huiyang@psu.edu; voice: 814-865-7397

## 1. Introduction

Nonlinear dynamics arise whenever multifarious entities of a system cooperate, compete, or interfere. Effective monitoring and control of nonlinear dynamics will increase system quality and integrity, thereby leading to significant economic and societal impacts. For example, heart disease is responsible for 1 in every 4 deaths in the United States, amounting to an annual loss of $448.5 billion [1]. Realizing a better quality of cardiac operations will reduce healthcare costs and improve the health of our society. Fig.1a shows nonlinear waveforms of 1-lead electrocardiogram (ECG) signals when human heart maintains blood circulation through *orchestrated depolarization and repolarization* of cells. It is common to observe the near-periodic patterns but with hidden temporal variations between heart cycles in these physiological signals. Fig 1a shows some common characteristics of ECG signals: 1) Within one cycle, the signal waveforms at different segments change significantly. The reason is that different segments often correspond to different stages of cardiac operations. 2) Between cycles, the signals are similar to each other but with variations. Near-periodical beatings of human heart provide nourishments to all parts of body and maintain vital living organs. 0

As complex physiological systems evolve in time, dynamics deal with change. Whether the system settles down to the steady state, undergoes incipient changes, or deviates into more complicated variations, it is dynamics that help analyze system behaviors. Fig. 1b shows an example of the ECG phase space constructed from multi-lead ECG signals using the Takens' embedding theorem [2]. As multiple sensors are deployed at various locations, distributed sensing provides multi-directional views of nonlinear dynamics in the underlying processes. Traditional linear methodologies focus on the analysis of *time-domain signals*, and attempt to understand a system's behavior by breaking it down into parts and then combining all constituent parts that have been examined separately. This idea underlies such methods as principal component analysis (PCA), Fourier analysis, and factor analysis. These methods encounter difficulties in capturing nonlinear, nonstationary and high-order variations. The breakthrough in nonlinear theory came with Poincaré's geometric thinking of dynamical systems [3, 4], *which focuses on geometric analysis of nonlinear trajectories in the phase space* (see Fig. 1b).

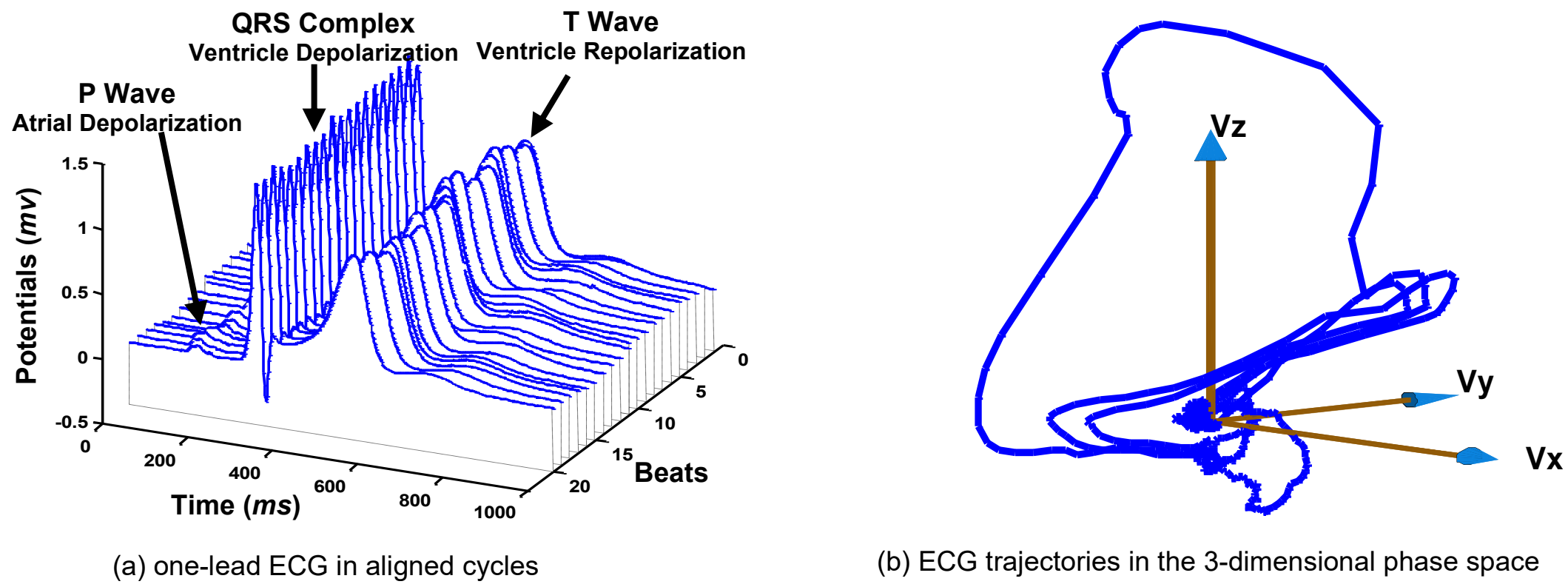

(a) one-lead ECG in aligned cycles

(b) ECG trajectories in the 3-dimensional phase space

**Fig. 1:** Examples of physiological signals: (a) aligned ECG cycles, (b) ECG trajectories.

Physiological sensing brings the proliferation of measurements of process dynamics (e.g., action potentials, ECG signals, echocardiogram). The challenge now is to harness and exploit nonlinear complexity underlying sensing signals for quality and integrity improvements in cardiac operations. However, multi-sensing capabilities are not fully utilized to extract information about nonlinear dynamics in the *phase-space domain*. Particularly, nonlinear dynamical systems defy understanding based on the traditional reductionist's approach, in which one attempts to understand a system's behavior by combining all constituent parts that have been analyzed separately. For example, clinicians had thought that drugs that significantly reduce arrhythmic behaviors in isolated cardiac cells would also do so in the heart until the concept was proven wrong by the failure of two large clinical trials [5]. In order to cope with system complexity and increase information visibility, modern healthcare systems are investing in advanced physiological sensing and patient monitoring, thereby giving rise to big data. Realizing the full potential of big data for healthcare intelligence requires fundamentally new methodologies to harness and exploit complexity. However, available nonlinear dynamics techniques are either not concerned with healthcare objectives or fail to effectively analyze big data to extract useful information for improving healthcare services. There is an urgent need to develop analytical methodologies that fully utilizing nonlinear dynamics and chaos principles for advancing healthcare services with exceptional features such as personalization, responsiveness, and superior quality.

Over the past few decades, the theory of nonlinear dynamics has emerged as a powerful technique in the design of superconducting circuits [6], chatter control in mechanical systems [7], laser stabilization [8], precise fabrication of nanomaterials [9], as well as information security [10]. In addition, several investigations into characterization and modeling of physiological systems, from the cellular level to the system level, have begun to adapt nonlinear dynamics and chaos principles. This chapter reviews some theoretical developments and tools to advance the applications of nonlinear dynamics principles in health care. Specifically, we will focus on the authors' recent investigations into sensor-based characterization and modeling of nonlinear dynamics in physiological systems. Case studies and applications in the studies of heart rate variability and space-time ECG signals will be presented. We hope that our limited and focused review will spur further development of nonlinear dynamics methodologies for improving healthcare services and accelerating the discovery of scientific knowledge in biomedical research.

The remainder of this chapter is organized as follows: Section 2 presents a primer on basic concepts of nonlinear dynamics and chaos. Section 3 will present two methods (i.e., multifractal analysis and multiscale recurrence quantification) for sensor-based characterization and modeling of nonlinear dynamics. Section 4 provides the case studies that adapts nonlinear dynamical systems principles for healthcare applications. Section 5 presents the discussion and conclusions arising out of this study.

## 2. Background

Nonlinear dynamics theory has emerged as an important methodology for complex systems modeling and analysis. The basic idea is to model the state evolution of underlying processes by a set of nonlinear differential equations, i.e., $\dot{\boldsymbol{X}} = \frac{dX}{dt} = F(\boldsymbol{X}, \boldsymbol{\theta}), F \in \mathbb{R}^n \to \mathbb{R}^n$, where $\boldsymbol{X}$ is a multi-dimensional state variable, $F$ is the nonlinear function, and $\boldsymbol{\theta}$ is model parameters. Thus, the solution, i.e., $\boldsymbol{X} = f(\boldsymbol{X}(0), t)$, generates a trajectory representing the flow of state evolution for a given initial condition $\boldsymbol{X}(0)$. When there is a small perturbation in $\boldsymbol{\theta}$ or $\boldsymbol{X}(0)$, the dynamics of a nonlinear process undergo abrupt changes and reveal complex characteristics, including chaos, recurrences, fractals and bifurcations. Notably, linear systems often attribute irregular behaviors of the system to random external inputs, but nonlinear systems can produce very chaotic data with purely deterministic equations and without stochastic inputs.

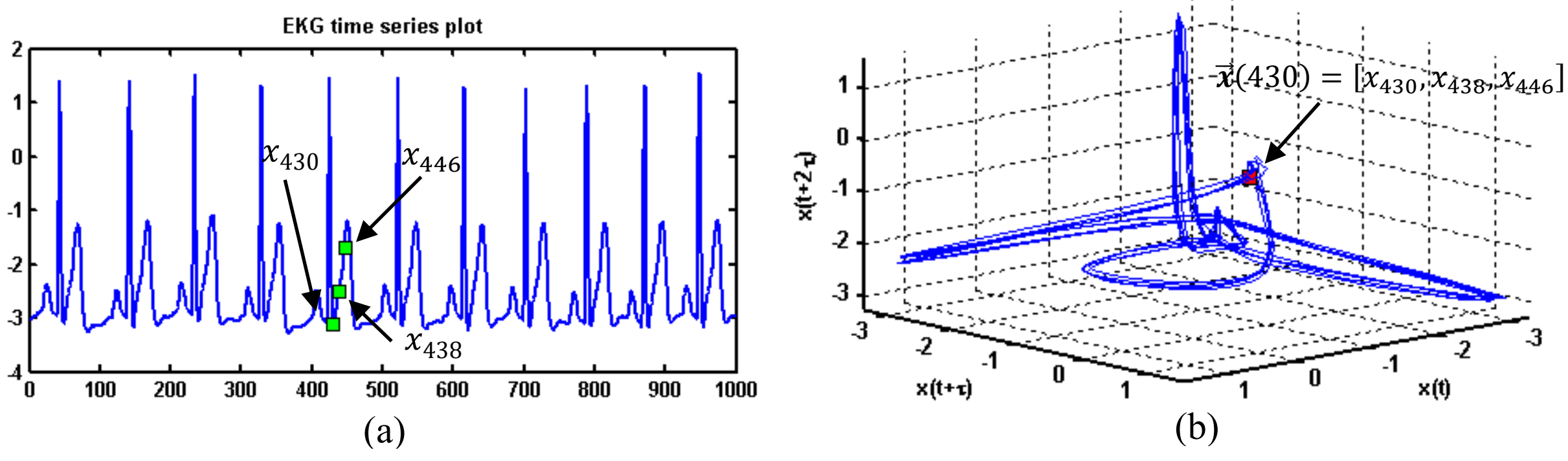

**Fig. 2:** An example of time delay reconstruction: (a) ECG time series, (b) lag-reconstructed ECG attractor.

Much of the complexity in real-world systems is known to emerge from the underlying nonlinear stochastic dynamics. The exhibited signals from complex systems are often chaotic in nature with irregular behaviors. However, dynamics manifest in the vicinity an attractor **A** (e.g., ECG attractor shown in Fig. 1b), an invariant set defined in an *m*-dimensional state space. Takens' delay embedding theorem [11] shows that system dynamics can be adequately reconstructed by using the time-delay coordinates of the individual measurements because of the high dynamic coupling existing in physical systems. For the time series $\boldsymbol{X} = \{x_1, x_2, \cdots, x_N\}^T$, state vector $\vec{\boldsymbol{x}}$ (Fig. 2b) is reconstructed using a delay sequence of $\{x_i\}$ as $\vec{\boldsymbol{x}}(i) = [x_i, x_{i+\tau}, \cdots, x_{i+\tau(m-1)}]$, where $m$ is the embedding dimension and $\tau$ is the time delay. Fig. 2 shows an example of time delay reconstruction of 3-dimensional ECG state attractor from the 1-dimensional ECG time series. The optimal embedding dimension $m$ suffice to unfold the attractor is determined by false nearest neighbor method [12]. In addition, mutual information [13] is used to minimize both linear and nonlinear correlations for the choice of optimal time delay $\tau$.

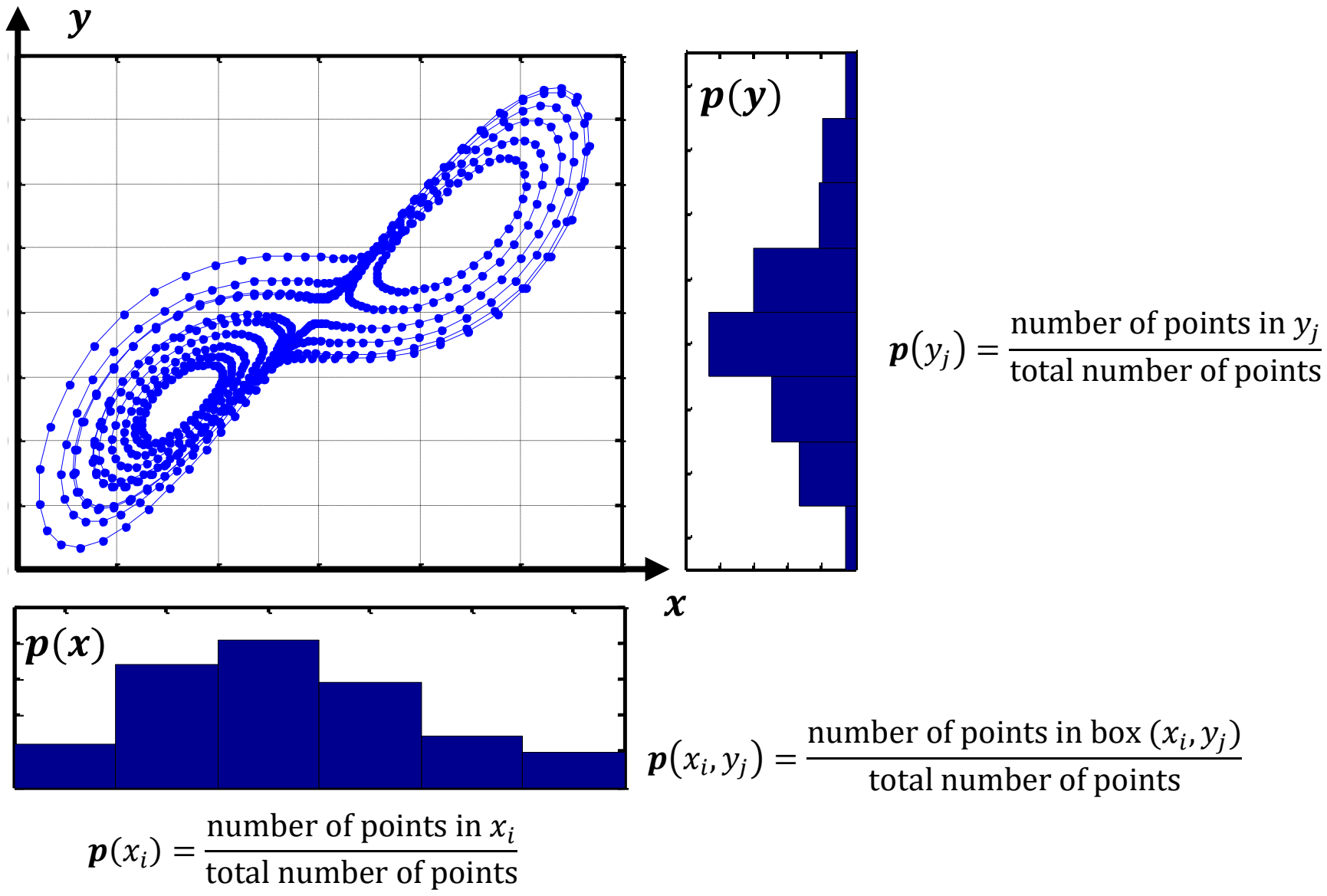

**Fig. 3**. An illustration for the computation of mutual information.

If the time delay $\tau$ is too small, the attractor will be *restricted to the diagonal* of the reconstructed phase space. However, if the time delay is too large, reconstructed attractor no longer represents the true dynamics. In the literature, there are two traditional approaches for the selection of time delay $\tau$. The first approach is to increase the $\tau$ value and then visually inspect that which $\tau$ gives the most spread out attractor. The disadvantage of visual inspection is that it only achieves satisfactory results for simple systems. The second approach is autocorrelation function (delay $\tau$)

$$r_\tau = \frac{\sum_{i=1}^{N-\tau}(x_i - \bar{x})(x_{i+\tau} - \bar{x})}{\sum_{i=1}^{N}(x_i - \bar{x})^2}$$

Optimal $\tau$ is required to minimize the linear independence that is the value when the autocorrelation function first passes through 0. Yet, autocorrelation is a second-order quantity evaluating merely linear dependency among data. Notably, mutual information quantifies both linear and nonlinear dependency between two variables $x_i$ and $y_j$, which is defined as:

$$I(x, y) = \sum_{i,j} p(x_i, y_j) \, log \frac{p(x_i, y_j)}{p(x_i) p(y_j)}$$

where $\mathrm{p(x, y)}$ is the joint probabilistic distribution, $\mathrm{p(x)}$ and $\mathrm{p(y)}$ are marginal probabilities. Fig. 3 shows the practical implementation to compute the mutual information. In the scatter plot of two variables x and y, the histogram is shown for each variable. Marginal probabilities $p(x_i)$ and $p(y_j)$ are computed as the number of points in $x_i$ and $y_j$ divided by the total number of points in the 2-dimensional space. The joint probability $p(x_i, y_j)$ is computed as the number of points in box $(x_i, y_j)$ divided by the total number of points in the space. Optimal $\tau$ is selected to minimize the general dependency between variables that is the first local minimum of Mutual Information function.

The method of false nearest neighbor (FNN) was first proposed by Kennel et al. to determine the minimal embedding dimension $m$ suffice to reconstruct system dynamics [12]. In other words, FNN method is to reconstruct the abstractor in the $m$-dimensional space that preserves dynamical properties of complex systems in the original phase space. Most importantly, the minimal dimension needs to guarantee the diffeomorphism of reconstruction without any information being lost but without adding unnecessary information. Suppose a $m$-dimensional attractor is projected to the lower dimensional space ($m'$ dimension and $m' < m$). Due to this projection, the topological structure of the $m$-dimensional attractor is no longer preserved. Some states are projected into neighborhoods of other states, but they are not true neighbors in the higher dimensional space. These states are called "false neighbors". An optimal dimension for time-delayed embedding is the smallest dimension that minimizes the number of "false neighbors". However, a larger dimension than the optimum leads to excessive computation when investigating the dynamical properties. "Noise" will populate and dominate the extra dimension of the space where no dynamics is operating. The basic idea of FNN is to measure the distances between a state and its nearest neighbors as this dimension increases. This distance should not change if the states are really nearest neighbors.

For a given time series $\boldsymbol{X} = \{x_1, x_2, \cdots, x_N\}^T$, we calculate the change of distances between neighboring states when the embedding dimension is increased from $\mathrm{m}$ to $\mathrm{m} + 1$. If the embedding dimension is high enough, then the fraction of false neighbors is zero, or at least sufficiently small. The state vector in m-dimensional space is

$$\boldsymbol{x}(i) = \left(x_i, x_{i+\tau}, \cdots, x_{i+\tau(m-1)}\right)$$

Let's denote the rth nearest neighbor of $\boldsymbol{x}(i)$ by $\boldsymbol{x}^{(r)}(i)$, then the Euclidean distance between $\boldsymbol{x}(i)$ and its neighbor is

$$R_m^2(\mathrm{i}, \mathrm{r}) = \sum_{k=0}^{m-1} \left(x_{i+k\tau} - x_{i+k\tau}^{(r)}\right)^2$$

If the embedding dimension is increased from $\mathrm{m}$ to $\mathrm{m} + 1$, the $(\mathrm{m} + 1)$th coordinate is added to each state vector $\boldsymbol{x}(i)$. Therefore, the distance between $\boldsymbol{x}(i)$ and the rth nearest neighbor that we identified in the mth dimension is

$$R_{m+1}^2(\mathrm{i}, \mathrm{r}) = R_m^2(\mathrm{i}, \mathrm{r}) + \left(x_{i+m\tau} - x_{i+m\tau}^{(r)}\right)^2$$

Then the FNN criterion (i.e., relative change in the distances between neighbors) is

$$\left(\frac{R_{m+1}^2(\mathrm{i}, \mathrm{r}) - R_m^2(\mathrm{i}, \mathrm{r})}{R_m^2(\mathrm{i}, \mathrm{r})}\right)^{1/2} = \frac{\left|x_{i+m\tau} - x_{i+m\tau}^{(r)}\right|}{R_m(i, r)} > R_{tol}$$

where $R_{tol}$ is the threshold. We will now examine the relative change in the distance as a way to see if the states are not really close together when increased to a higher-dimensional space.

## 3. Sensor-based characterization and modeling of nonlinear dynamics

Nonlinearity is one of the most ubiquitous properties of physiological systems. Sensor signals capture rich information on the underlying nonlinear dynamics in physiological processes. Linear systems often attribute irregular behaviors of the system to random external inputs, but nonlinear systems can produce chaotic data with purely deterministic equations and without stochastic inputs. Modeling and analysis of nonlinear systems are more challenging than those for a linear system. Effective strategies for modeling and monitoring of physiological systems need to consider enormous amount of sensing data as well as the nonlinear evolution of state variables in the underlying process. In this section, we will present a detailed review of two methodologies, namely multifractal analysis and multiscale recurrence quantification for sensor-based characterization and modeling of nonlinear dynamics.

### 3.1 Multifractal spectrum analysis of nonlinear time series

#### 3.2.1 Fractal dimension

The dimension is generally defined as the minimal number of coordinates one has to use to describe a point within the space. For example, a line needs one coordinate to specify a point on it and therefore its dimension is 1. Similarly, the dimension of a plane is 2 and the dimension of a cube is 3. The *topological dimension* of a set X takes integer values, and is defined by induction as 1 + the dimension of its boundary. In other words, the set X has a dimension of d if $\forall x \in X$, there is an arbitrarily small neighborhood of x whose boundary has a dimension of $d-1$. A set is zero dimensional if there is an arbitrarily small neighborhood of any point x whose boundary is empty. Because the notion of boundary is well defined in mathematics, the inductive dimension effectively describes topological spaces. Indeed, the topological dimension of $\mathbb{R}^d$ is d.

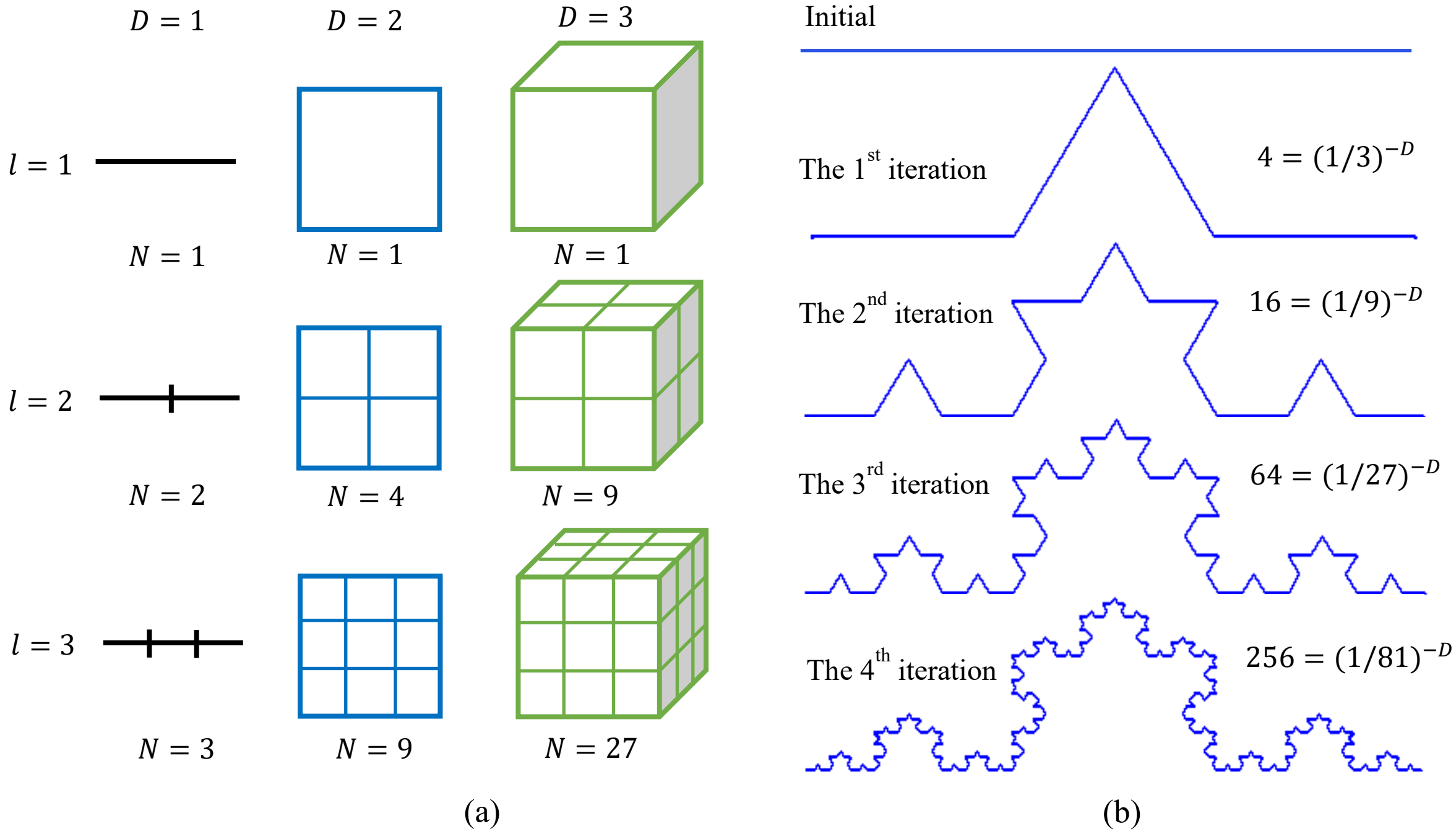

**Fig. 4**. Illustration of self-similarity and fractal dimension from the perspectives of scaling and covering for (a) Euclidean Geometry, i.e., line, square and cube and (b) Koch curve.

However, fractals are irregular geometric objects that cannot be sufficiently specified using topological dimensions. Fractal objects are self-similar, that is, look similar regardless of the magnification. If one zooms in or out the fractal set, its geometric shape has a similar appearance. Hence, fractal dimension is introduced to describe such "infinitely complex" fractal objects (or shape). Notably, fractal dimension is not topological. Fig. 4 illustrates the concepts of self-similarity and fractal dimension from the perspectives of scaling and covering. If we reduce the linear size of an object in the Euclidean space $\mathbb{R}^D$ by the scaling factor a in each spatial direction, its measure (length, area, or volume) will increase to $N = a^{-D}$, where $N$ is the number of measure elements to cover the object. As shown in Fig. 4a, if we reduce the size of a line by $a = 1/2$, then its measure (i.e., length) will increase to $N = (1/2)^{-1} = 2$. In other words, two measure elements are needed to cover the original line. Further, if we reduce the size of a line by 1/3, then its measure will be $N =$ $(1/3)^{-1} = 3$. However, if we reduce the size of a square by 1/2, then its measure (i.e., area) will increase to $\mathrm{N} = (1/2)^{-2} = 4$. In addition, nine measure elements ($\mathrm{N} = (1/3)^{-2}$) are needed to cover the original square if the line size of the square is reduced by 1/3 in each spatial direction. The scaling rule also holds for the cube. If we reduce the line size of a cube by 1/2 in each spatial direction, then its measure (i.e., volume) will increase to $\mathrm{N} = (1/2)^{-3} = 8$. If we reduce the size of a square by 1/3, then its measure will be $\mathrm{N} = (1/3)^{-3} = 27$. Fig. 4a shows how the measure changes with respect to linear scaling. If we take log of both sides of the relationship $\mathrm{N} = a^{-D}$, the dimension is $\mathrm{D} = -\log N/\log a$.

The dimension D needs not to be an integer, as shown for Euclidean geometry in Fig. 4a. Fig. 4b shows an example of the Koch snowflake curve which has a non-integer dimension. The Koch snowflake curve is generated by starting with a straight line, divide the line into three segments of equal lengths, and then remove the middle third of the line and replace it with two lines that have the same length (1/3) as the remaining lines in both sides. This process recursively iterates to generate the "infinitely complex" Koch curve. Fig. 4b shows the first 4 iterations of the process. In each iteration, the length of the curve increases. However, the Koch snowflake curve is self-similar at all scales of magnification. If we follow the scaling and covering rule, the dimension of Koch curve is $\mathrm{D} = -\log 4/\log(\frac{1}{3}) = 1.26$.

Fractal sets have theoretical dimensions that exceed their topological dimensions and can be non-integer values. Self-similarity across scales is a typical characteristic of fractals. Fractal dimension specifies the complexity of a fractal object by measuring the changes of coverings relative to the scaling factor. It also characterizes the space-filling capacity of a fractal object with respect to its scaling properties in the space. Many real-world objects exhibit self-similarity, e.g., scribbles, dust, ocean waves, or clouds. In practice, the relationship between scaling and covering is often difficult to be determined. The box-counting method is widely used to estimate the fractal dimension of an irregular object. The basic idea is to cover a fractal set with measure elements (e.g., box) at different scales and examine how the number of boxes changes with respect to the scaling factor [14, 15]. If N(a) is the number of boxes that are needed to cover a fractal object at the scale a, then the fractal dimension $D_B$ specifies how N(a) varies with respect to the scaling factor a as:

$$\mathrm{N(a)} \propto (1/a)^{D_B}$$

In general, the box-counting dimension is defined as

$$D_B \coloneqq \lim_{a \to 0} \frac{\ln N(a)}{\ln(1/a)}$$

However, the box-counting dimension $D_B$ may not exist if the limit does not exist. As the upper and lower limit always exist, the upper and lower bounds of box-counting dimension will be:

$$\overline{D}_B = \lim_{a \to 0} \sup \frac{\ln N(a)}{\ln(1/a)}, \underline{D_B} = \lim_{a \to 0} \inf \frac{\ln N(a)}{\ln(1/a)}$$

The box-counting dimension $D_B$ is well-defined when the two bounds are sufficiently close to each other.

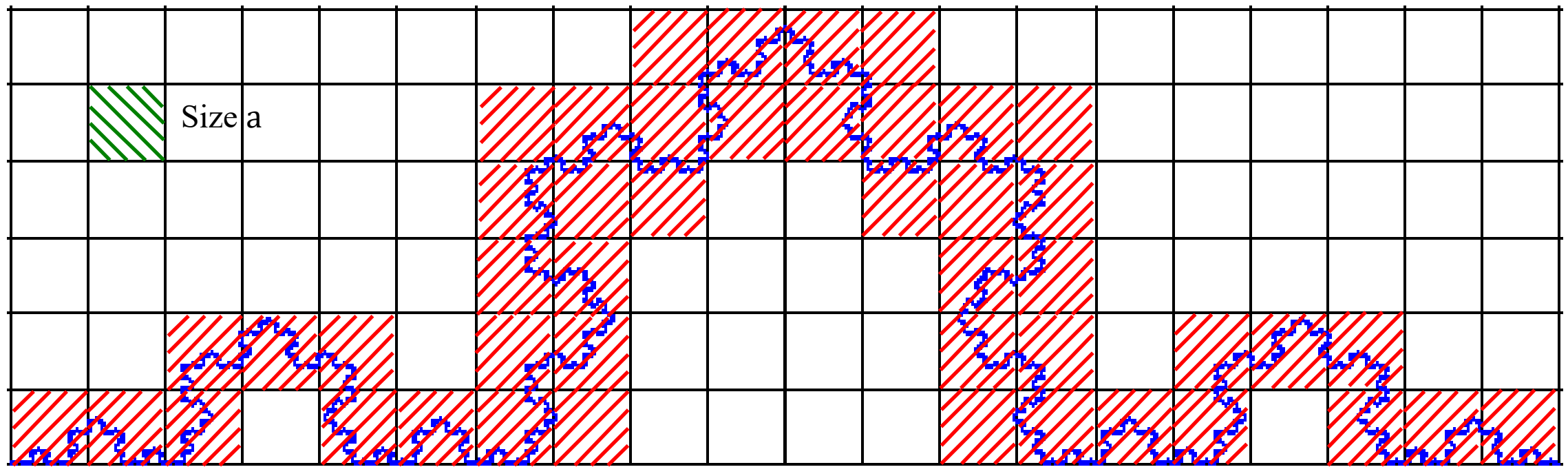

**Fig. 5**. An illustration of box counting method to cover the Koch curve with the box of size a.

Fig. 5 illustrates the use of box-counting method to calculate the fractal dimension of the Koch curve. The number of boxes $\mathrm{N(a)}$ required to cover the Koch curve increases when the "box" size decreases, and their relationship follows the power law, i.e., $\ln \mathrm{N(a)}$ is proportional to $\ln(1/a)$. If $\mathrm{N(a)}$ is computed for a range of a, there is a linear relationship between $\ln\mathrm{N(a)}$ and $\ln(1/a)$. The slope is an estimate of the fractal dimension. Fig. 6 shows the plot of $\ln\mathrm{N(a)}$ against $\ln(1/a)$ for the Koch curve. It is shown that $\ln\mathrm{N(a)}$ linearly increases with respect to $\ln(1/a)$ and the estimated slope is approximately 1.2849, while the theoretical fractal dimension of Koch curve is $\ln 4/\ln 3 = 1.262$.

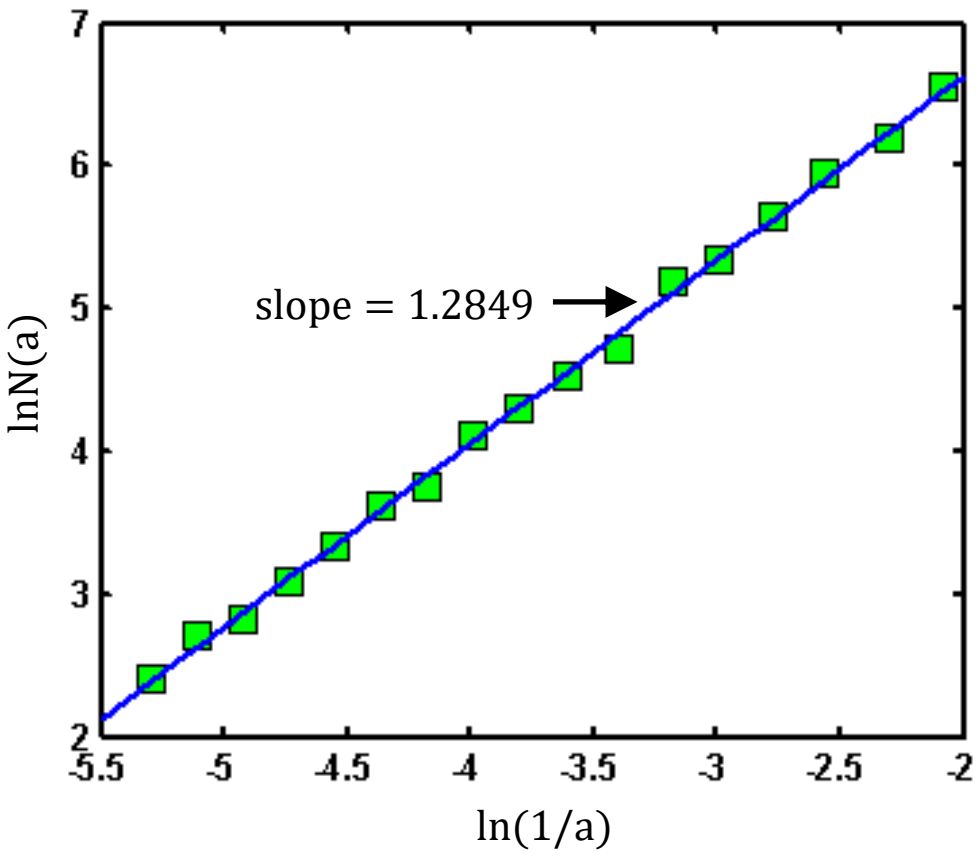

**Fig. 6**. Plot of $\ln(1/a)$ against $\ln\mathrm{N(a)}$ for Koch curve, the slope is the estimated fractal dimension.

However, there are several drawbacks in the box-counting method when estimating the dimension of a fractal set. First, if the upper and lower bounds of box-counting dimension are not close to each other, then $D_B$ is not well-defined. Second, the upper bound $\overline{D}_B$ may not be countably stable, i.e., $\overline{D}_B(\bigcup_{i=1}^{\infty} X_i) \neq \sup_i\{\overline{D}_B(X_i)\}$, where $X_i$ is the subset of a fractal set $\mathrm{X} = \bigcup_{i=1}^{\infty} X_i$. Third, the lower bound $\underline{D}_B$ may not be finitely stable, i.e., $\underline{D}_B(X_i \cup X_j) \neq \max(\underline{D}_B(X_i), \underline{D}_B(X_j))$, where $\mathrm{i} \neq \mathrm{j}$. Therefore, Hausdorff dimension $D_H$ is further introduced to characterize the fractal set $\mathrm{X} \subseteq \mathbb{R}^D$ [14]. For $\varepsilon > 0$, an $\varepsilon$-cover of X is a finite or countable collection of $\{\mathrm{B_i}\}_{\mathrm{i}=1,2,\ldots}$, where the ball $\mathrm{B_i} \subseteq \mathbb{R}^D$ and its diameter $|\mathrm{B_i}|$ is less than or equal to $\varepsilon$. The $\delta$-total length of $\{\mathrm{B_i}\}_{\mathrm{i}=1,2,\ldots}$ is defined as $\sum_{i=1}^{\infty}|\mathrm{B_i}|^{\delta}$. If $\{\mathrm{B_i}\}_{\mathrm{i}=1,2,\ldots}$ is a countable cover of the fractal set X, then the δ-dimenional Hausdorff mesure of X is defined to be the limit of the infimum of the $\delta$-total length of $\{\mathrm{B_i}\}_{\mathrm{i}=1,2,\ldots}$.

$$H^{\delta}(\mathrm{X}) = \lim_{\varepsilon\to 0} \inf\left\{\sum_{i=1}^{\infty}|\mathrm{B_i}|^{\delta} : \{\mathrm{B_i}\}_{\mathrm{i}=1,2,\ldots} \text{ is the } \varepsilon - \text{cover of X}\right\}$$

The Hausdorff dimension $D_H$ of the fractal set X exceeds its topological dimension and is defined as

$$H^{\delta}(\mathrm{X}) = \begin{cases} \infty \text{ if } \delta < D_H \\ 0 \text{ if } \delta > D_H \end{cases}$$

Furthermore, it is worth mentioning that monofractal analysis (i.e., a single fractal dimension) often fails to fully characterize complex scaling behaviors of many irregular objects in the real world [16]. Instead, multifractal analysis utilizes a spectrum of singularity exponents to provide a detailed description of complex scaling behaviors. Let's denote μ as a measure using the ball $\mathrm{B_i(a)}$ centered at $x_i$ of the object. Then the singularity exponent h at location $x_i$ will be

$$h(x_i) = \lim_{a\to 0^+} \frac{\ln\mu(\mathrm{B_i(a)})}{\ln(1/a)}$$

The singularity spectrum $\mathrm{D(h)}$ is the fractal dimension of the set of all the locations x such that $\mathrm{h}(x) = \mathrm{h}$:

$$\mathrm{D(h)} = \mathrm{D}_F(\{x\text{: h(x)} = \mathrm{h}\})$$

where $\mathrm{D}_F$ is the fractal dimension. The singularity spectrum D(h) provides a statistical distribution of singularity exponents h(x). Fig. 7a shows an example of the multifractal set, namely triadic Cantor set. The Cantor set is constructed as follows:

1. At step k = 0, the weight $\mu_0 = 1$ is assigned to the interval [0,1].
2. At step k = 1, the whole interval is divided into three subintervals of equally lengths. The new weights will be: $\mu_1 = p_1\mu_0 = p_1$ for the first subinterval [0,1/3] and $\mu_2 = p_2\mu_0 = p_2$ for the third subinterval [2/3,1], where $p_1$ and $p_2$ are two probability values. The second subinterval will have a zero weight.
3. This process is iteratively repeated and then weights are summed over all the steps in the interval [0,1] to generate the Cantor set.

After k steps, if we consider the first interval $\mathrm{B}_1(\mathrm{a} = 3^{-k}) = [0, 3^{-k}]$, then the measure $\mu(\mathrm{B}_1) = p_1^k\mu_0 = p_1^k$ at the location $x = 0$. Thus, the singularity exponent at $x = 0$ is $\mathrm{h}(x = 0) = -\ln p_1/\ln 3$. Similarly, one can prove that the singularity exponent at $x = 1$ is $h(x = 1) = -\ln p_2/\ln 3$ for the last interval $\mathrm{B}_{2^k}(\mathrm{a} = 3^{-k}) = [1 - 3^{-k}, 1]$. If $p_1 = p_2$, then we will have a monofractal Cantor set. If $p_1 \neq p_2$, then there will be a spectrum of singularity exponents, particularly $\mathrm{h}(x = 0) \neq \mathrm{h}(x = 1)$. Fig. 7a illustrates a multifractal version of triadic cantor set with $p_1 = 0.6$ and $p_2 = 0.4$. Moreover, the singularity exponents are $h(\mathrm{x} = 0) = 0.834$ and $h(\mathrm{x} = 1) = 0.465$. As such, we will have $\mathrm{D(h = 0.834)} = \mathrm{D}_F(\{x = 0\}) = 0$ and $\mathrm{D(h = 0.465)} = \mathrm{D}_F(\{x = 1\}) = 0$. Fig. 7b shows the singularity spectrum D(h). Instead of a sole singularity exponent, there is a range of singularity exponents that describes complex scaling behaviors of multifractal Cantor set.

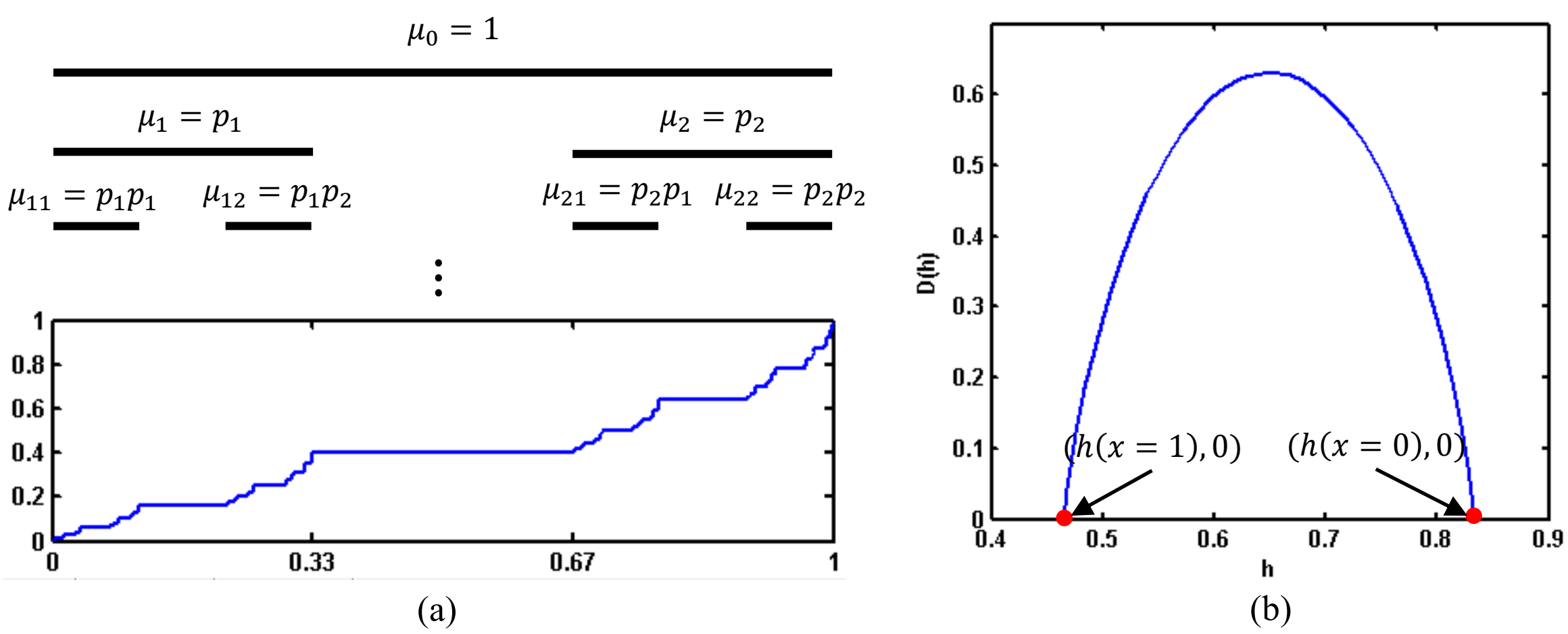

**Fig. 7**. (a) Triadic Cantor set with $p_1 = 0.6$ and $p_2 = 0.4$ and (b) D(h) singularity spectrum.

### 3.2.2 Continuous wavelet transformation

Traditionally, box-counting methods leverage the measure elements (e.g., box, square, line) at different spatial scales to cover the fractal set and then examine how the covering changes with respect to the scaling factor. Here, the fractal set refers to an irregular object in the space. The scaling factor refers to the variations of spatial scales of measure elements. However, box-counting methods are not generally applicable to measure fractal dimension of complex time series. First, the scaling factor usually refers to temporal scales instead of spatial ones for a time series. Second, time series involves a range of frequency components that are not specifically considered when dealing with an irregular object in the space. Third, the box-counting technique cannot adequately address the challenge of low-frequency trends (e.g., polynomial) in the time series and thereby fail to measure local scaling properties. New methods and tools to characterize scaling behaviors and quantify fractal dimensions of time series are urgently needed.

Therefore, wavelet functions are widely used as "boxes" in multifractal spectrum analysis of time series. Wavelet functions are building blocks that can be used to simultaneously decompose signal characteristics in both time and frequency domains. Wavelet representations delineate steady and transient components of nonstationary time series into various frequency bands while preserving the time information. In particular, wavelet transform effectively addresses polynomial trends that fail the traditional box-counting techniques. Time-frequency representation is particularly useful for revealing the underlying hierarchy that governs the temporal distribution of local singularity exponents.

The continuous wavelet transform (CWT) is an effective time-frequency representation that overcomes the resolution problems in the short time Fourier transform (STFT) [17]. Notably, Fourier analysis interprets the regular structure, e.g., dominant frequencies in the signals, but does not provide the temporal localization of frequency components and assume spectral components exist at all times (i.e., stationarity). Therefore, STFT employs a local analysis scheme for a time-frequency representation (TFR) of nonstationary signals. STFT segments the time series into narrow time windows, narrow enough to be considered stationary, and then take the Fourier transform of each segment. Further, CWT uses a variable-length wavelet function to address the preset resolution problem of STFT. As shown in Fig. 8, a narrower wavelet function captures high-frequency transient behaviors in a fine-grained time resolution, and the wider one characterizes low-frequency steady behaviors in a better frequency resolution.

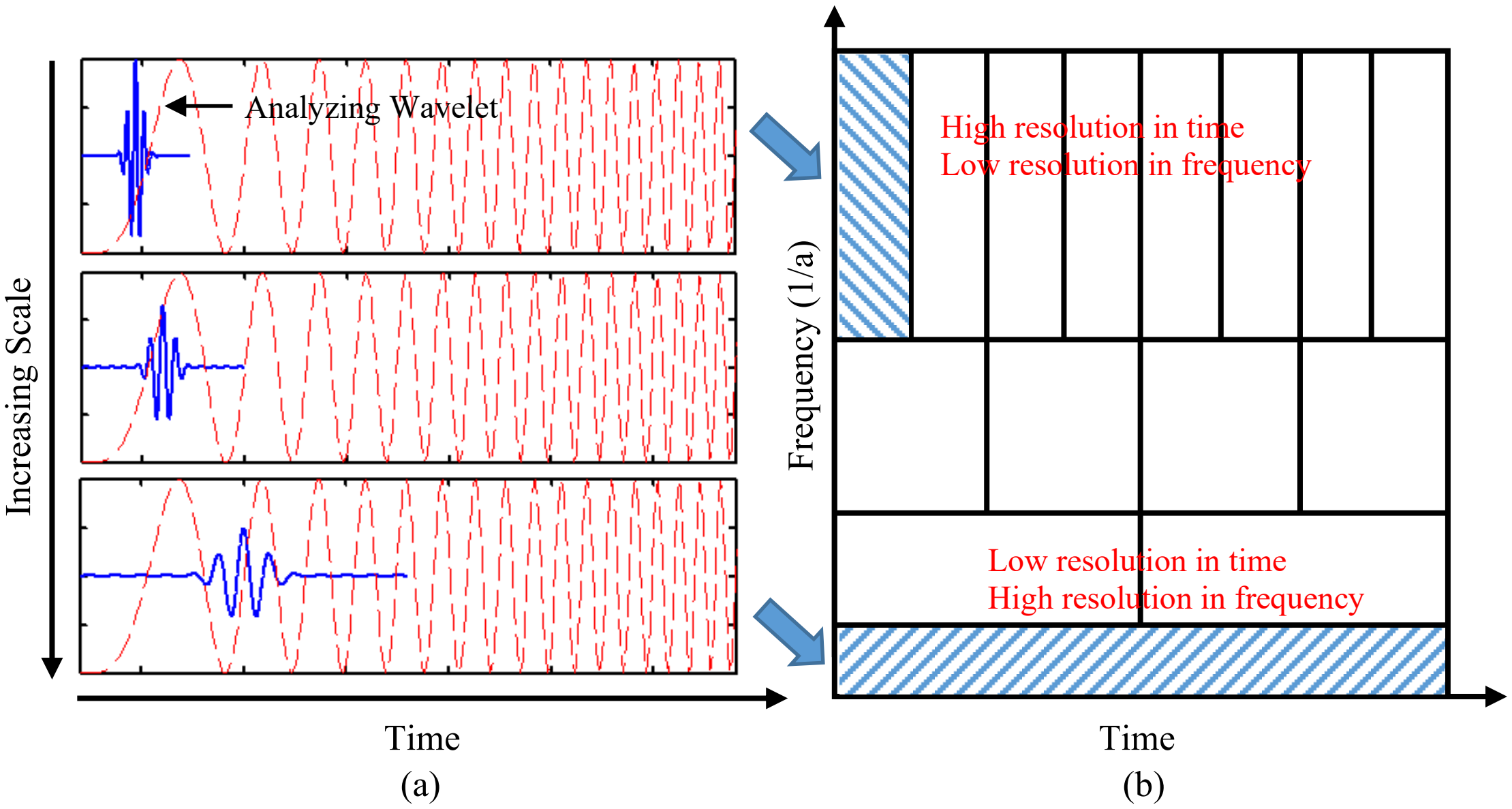

**Fig. 8**. (a) Continuous wavelet transform of time series with an analyzing wavelet at different scales and time locations. (b) Time-frequency resolution of wavelet representation.

The continuous wavelet transform of signal $x(t)$ using the analyzing wavelet $\psi(\cdot)$ is defined as

$$CWT_x^{\psi}(b,a) = \Psi_x^{\psi}(b,a) = \frac{1}{\sqrt{|a|}}\int_t x(t)\psi^*\left(\frac{t-b}{a}\right)dt$$

where $CWT_x^{\psi}$ represents wavelet coefficients, a is the scale parameter (i.e., measure of frequency) and b is the translation parameter (i.e., measure of time). The mother wavelet $\psi(\cdot)$ is translated and scales to obtain all kernels $\psi(\frac{t-b}{a})$. In other words, $CWT_x^{\psi}(b,a)$ is the cross correlation of the signal $x(t)$ with the mother wavelet at the scale $a$ and at the time lag of $b$. If $x(t)$ shares similar patterns to the wavelet function $\psi(\frac{t-b}{a})$ at the time location b, then wavelet coefficients $CWT_x^{\psi}(b,a)$ will be large.

As shown in Fig. 8a, the mother wavelet is shifted in the time domain with the translation parameter as $\psi(t-b)$ and is expanded or compressed with the scale parameter as $\psi(\frac{t}{a})$. Because scale is inversely proportional to frequency, smaller scales are corresponding to more compact wavelet functions (i.e., high frequency). Continuous wavelet transform starts with the small-scale wavelet functions (high frequency) and then proceeds to larger-scale ones (low frequency) where the wavelet function is more expanded. The wavelet function is first set at the beginning of the signal. The inner product of the signal and wavelet function is then computed. The results are normalized by the factor $1/\sqrt{|a|}$, which is ensure that wavelet functions have the same energy. The wavelet function is shifted along the time direction and new coefficients will be calculated. This process is continued until reaching the end of the signal.

The variable-length wavelet function is analogues to flexible windowing in the continuous wavelet transform. Hence, wavelet representation provides a better time-frequency resolution as demonstrated in Fig. 8b. As aforementioned, small-scale wavelets are more compact (i.e., similar to the use of small window in STFT) and therefore capture the high-frequency components in the time series. Large-scale wavelets are more stretched and have a bigger window size, thereby capturing the low-frequency components. As shown in Fig. 8b, when the window size is smaller, time resolution is higher but frequency resolution is lower. However, when the window size is bigger, time resolution is lower but frequency resolution is higher. The use of variable-length wavelet functions in CWT provides a trade-off between frequency and time resolutions [17, 18], which is advantageous in multifractal spectrum analysis of complex time series that will be detailed in the next section.

### 3.2.3 Multifractal spectrum analysis

The wavelet transform modulus maxima (WTMM) method is widely used to quantify multifractal spectrum of a nonlinear time series. This wavelet-based multifractal analysis evaluates the local singularity exponent $h$ through the continuous wavelet transform. Note that the WTMM method uses wavelets in different scales as the box functions to measure the self-similarity in the time-frequency representation of time series [19, 20]. Suppose μ is a measure of a fractal set, $\{\mathrm{B_i(a)}\}_{i=1,2,\cdots,N(a)}$ is a covering of the support of μ, where $\mathrm{B_i(a)}$ is the $ith$ box of size $a$ and $\mathrm{N(a)}$ is the number of boxes. For $q \in \mathbb{R}$, the partition function $\mathrm{Z(q, a)}$ is defined as

$$Z(q,a) = \sum_{i=1}^{N(a)} \mu_i^q(a)\,, \text{where } \mu_i(a) = \mu(\mathrm{B_i(a)}) = \int_{\mathrm{B_i(a)}} d\mu$$

Because the rigid box function leads to smooth behaviors that will distort the singularities of time series and impair the estimation of local singularity exponent, wavelet functions are used to substitute traditional box functions. Therefore, the partition function in the new wavelet multifractal formalism is

$$Z(q,a) = \int \left|\Psi_x^{\psi}(b,a)\right|^q db$$

where $\Psi_x^{\psi}(b,a)$ are wavelet coefficients at location $b$ and scale a. In order to improve the estimability of singularity and avoid the divergence of $Z(q,a)$ for $\mathrm{q} < 0$, the integration is further modified to be a discrete summation over the maxima of $\Psi_x^{\psi}(b,a)$. Hence, the partition function is revised as

$$Z(q,a) = \sum_{l \in \mathcal{L}(a)} \left|\Psi_x^{\psi}(b_l(a),a)\right|^q$$

where $l \in \mathcal{L}(a)$ denotes the maxima line at the scale a, and $b_l(\mathrm{a})$ is the position of the maxima belonging to the line $l$ at the scale a. Fig. 10b shows an example of continuous wavelet transform and the black lines are the maxima of wavelet coefficients. More details will be given in the later examples.

Furthermore, because the maxima line is sparse in the wavelet representation and the partition function is unstable for the negative values of q, the WTMM method defines the partition function by replacing the wavelet transform modulus maxima at the scale a by the supremum values along the maxima line at scales smaller than a.

$$Z(q,a)=\sum_{l\in\mathcal{L}(a)}\left(\sup_{a'\leq a}\left|\Psi_x^{\psi}(b_l(a'),a')\right|\right)^q$$

where $\Psi_x^{\psi}(b_l(a'),a')$ are wavelet transform coefficients at location $b_l(a')$ and scale $a'$, $\sup_{a'\leq a}|\cdot|$ is the local maxima of modulus for all scales $a'\leq a$, $l\in\mathcal{L}(a)$ denotes the maxima line at the scale a. Hence, $Z(q,a)$ is the sum of q-th powers of maxima's in wavelet modulus. When $a\to 0^+$, $Z(q,a)$ can be approximated as $Z(q,a)\cong a^{\tau(q)}$, where $\tau(q)$ is the spectrum of singularity exponents that describes the power law scaling behavior of $Z(q,a)$ with respect to the scale a. As mentioned in the section 3.2.1, $\mu_i(a)\propto a^{h(x_i)}$ and $N(a)\propto a^{-D}$. Therefore, the partition function can be approximately expressed as

$$Z(q,a)=\sum_{i=1}^{N(a)}\mu_i^q(a)\cong N(a)\cdot\mu^q(a)\propto\int a^{qh-D(h)}da$$

When $a\to 0^+$, this integration is dominated by the term $a^{qh-D(h)}$. Hence, we have $\tau(q)=qh(q)-D(h)$, where the local singularity exponent $h(q)$ is not constant and is calculated as $h(q)=d\tau(q)/dq$. Further, the multifractal spectrum $D(h)$ can be derived from $\tau(q)$ through a Legendre transform [21].

$$D(h)=qh-\tau(q)$$

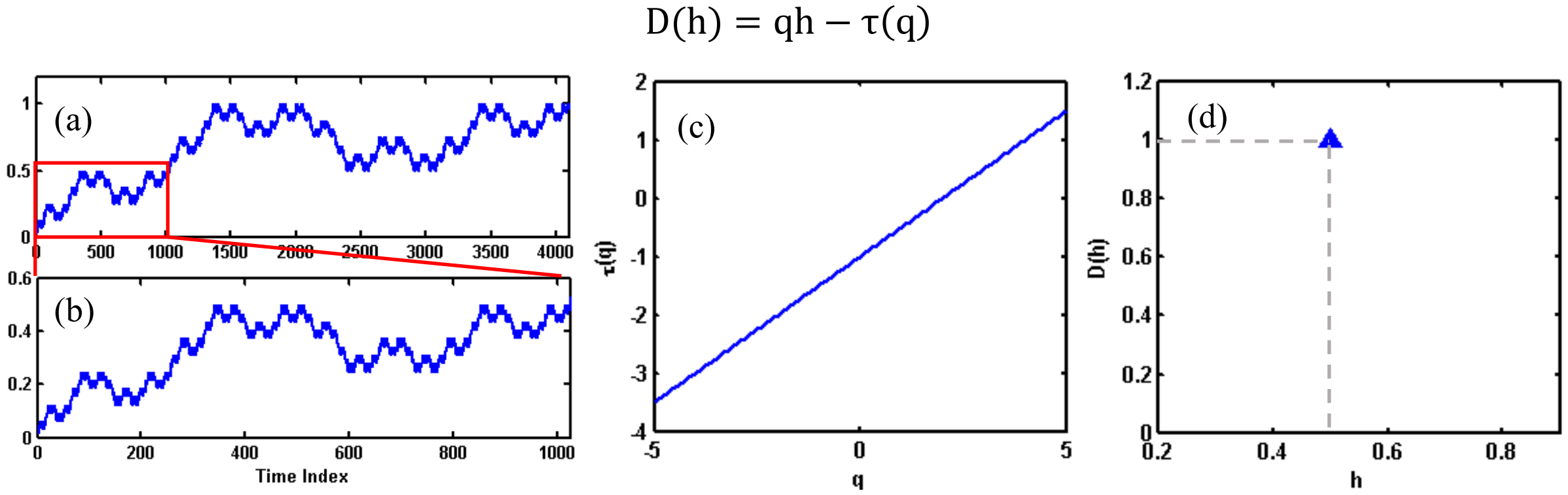

**Fig. 9:** (a) The generalized devil staircase with weights $p_1=p_2=-p_3=p_4=0.5$; (b) Singularity spectrum $\tau(q)$ versus $q$; (c) Multifractal spectrum $D(h)$ versus $h$.

Furthermore, we will introduce the generalized devil staircase [22] to demonstrate the differences between monofractal and multifractal sets and the use of WTMM method to derive the multifractal spectrum. The devil staircase is constructed as follows:

1. At the first step, if an interval is divided into four subintervals with equal lengths and different weights $p_1$, $p_2$, $p_3$ and $p_4$, then we will have the scale $a=4^{-1}$ and four measures $\mu_i=|p_i|$ ($i=1,2,3,4$). As such, the partition function is $Z(q,a)=\sum_{i=1}^{4}\mu_i^q(a)=|p_1|^q+|p_2|^q+|p_3|^q+|p_4|^q$.

2. If the process is iterated for each subinterval, then there will be $4^K$ measures (or subintervals) of size $a=4^{-K}$ to cover the whole interval. Each measure will have the form $\prod_{k=1}^{K}|\tilde{p}_k|$, where $\tilde{p}_k\in\{p_i,i=1,2,3,4\}$. The partition function is $Z(q,a)=(|p_1|^q+|p_2|^q+|p_3|^q+|p_4|^q)^K$.

3. Finally, the generalized devil staircase $x_{ds}(t)$ is the cumulative distribution function of all subintervals at the $K$th step.

Note that the constraint of weights is $\sum_{i=1}^{4}p_i=1$ and $p_i\in\mathbb{R},i=1,2,3,4$, which is designed to reach convergence. Base on the definition of $\tau(q)$, one can prove that the singularity spectrum $\tau(q)$ is

$$\tau(q) = \frac{\ln Z(q,a)}{\ln a} = \frac{\ln(|p_1|^q + |p_2|^q + |p_3|^q + |p_4|^q)^K}{\ln(1/4^K)} = -\log_4(|p_1|^q + |p_2|^q + |p_3|^q + |p_4|^q)$$

Fig. 9a shows the devil staircase with the weights $p_1 = p_2 = -p_3 = p_4 = 0.5$. Here, $x_{ds}(t)$ is generated after 6 iterations, and hence the fractal set includes $4^6 = 4096$ subintervals. At each iteration of the construction, the order of weights follows exactly $p_1, p_2, p_3$ and $p_4$. Note that the devil staircase $x_{ds}(t)$ is everywhere continuous but nowhere differentiable. If we zoom into the first quarter of the devil staircase $x_{ds}(t)$, Fig. 9b shows distinct self-similar patterns. In addition, we will have the singularity spectrum (see Fig. 9c) as

$$\tau(q) = -\ln_4(|p_1|^q + |p_2|^q + |p_3|^q + |p_4|^q) = -\ln_4\left(4\left(\frac{1}{2}\right)^q\right) = \frac{1}{2}q - 1$$

Therefore, the singularity exponent will be $h(q) = d\tau(q)/dq = 1/2$. The fractal dimension is $D(h) = hq(h) - \tau(q(h)) = \frac{1}{2}q - \left(\frac{1}{2}q - 1\right) = 1$ (see Fig. 9d). As a result, the devil staircase with weights $p_1 = p_2 = -p_3 = p_4 = 0.5$ is monofractal.

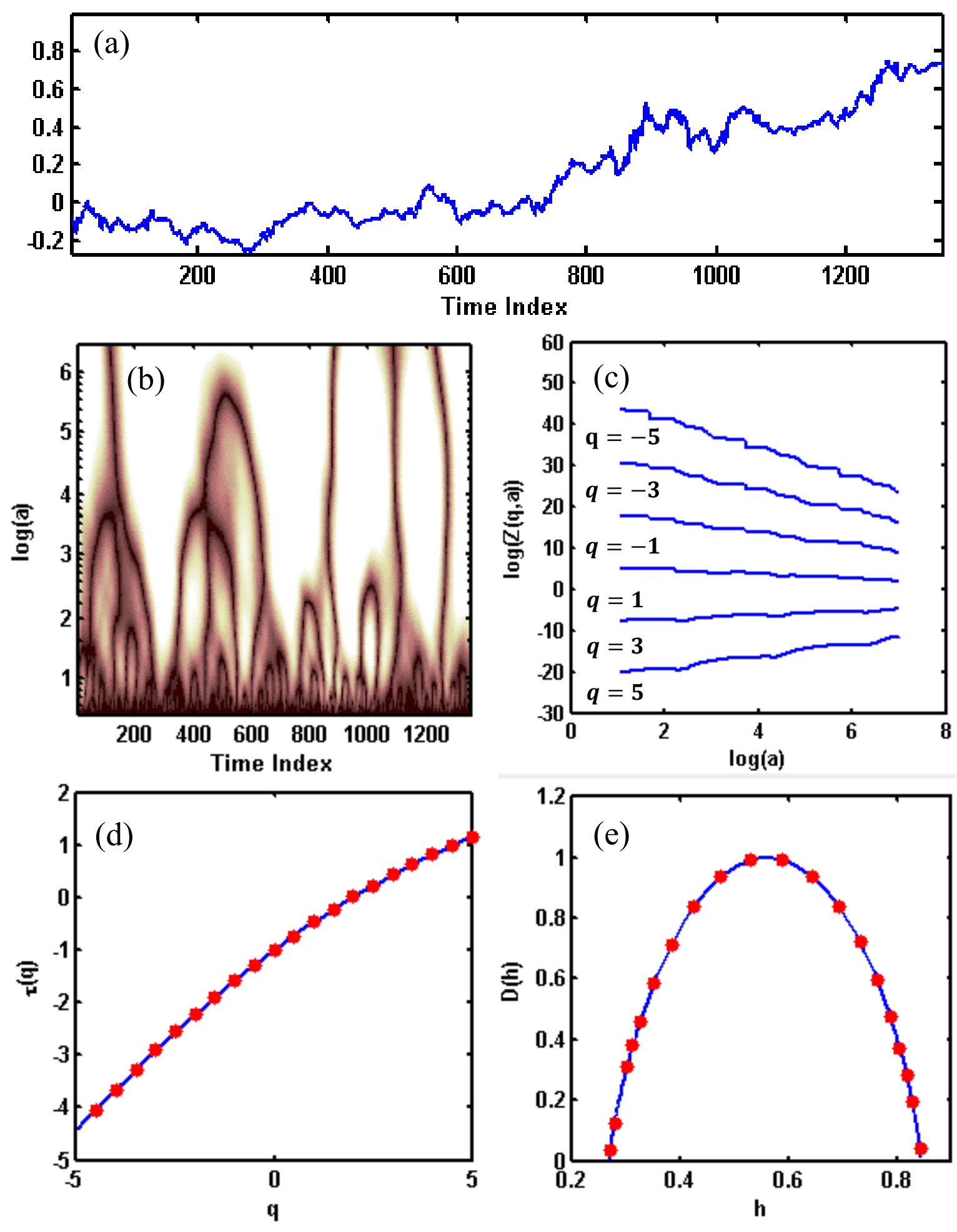

**Fig. 10:** (a) The generalized devil staircase with weights $p_1 = 0.69,\ p_2 = -p_3 = 0.46,\ p_4 = 0.31$; (b) Continuous wavelet transform; (c) $\log(Z(q,a))$ vs. $\log(a)$; (d) $\tau(q)$ vs. $q$; (e) $D(h)$ vs. $h$. Note: red dots are calculated from the WIMM method, and the blue lines corresponds to the theoretical curves.

Fig. 10a shows the random devil staircase with the weights as $p_1 = 0.69,\ p_2 = -p_3 = 0.46,\ p_4 = 0.31$, where the relation $p_1 + p_2 + p_3 + p_4 = 1$ still holds. However, at each iteration of construction, the order of weights is chosen randomly from $p_1, p_2, p_3$ and $p_4$. Here, $x_{ds}(t)$ is generated after 5 iterations, and hence the fractal set includes $4^5 = 1024$ subintervals. Due to the randomness in each iteration, the random devil staircase function $x_{ds}(\mathrm{t})$ and the singularity spectrum are difficult to be expressed analytically. Therefore, the WTMM method is used to quantify the multifractal spectrum of the random devil staircase $x_{ds}(\mathrm{t})$.

Fig. 10b shows the continuous wavelet transform of the random devil staircase $x_{ds}(\mathrm{t})$, where the black lines represent maxima lines. Fig. 10c shows $\log(\mathrm{Z(q,a)})$ versus $\log(\mathrm{a})$ for different q values, which is numerically calculated based on wavelet maxima modulus and the definition of partition function $Z(q,a) = \sum_{l\in\mathcal{L}(a)}\left(\sup_{a'\le a}\left|\Psi_x^{\psi}(b_l(a'),a')\right|\right)^q$. Because the singularity spectrum is $\tau(\mathrm{q}) = \frac{\ln Z(q,a)}{\ln a}$, the slope of each curve for different $\mathrm{q}'s$ in Fig. 10c will be its corresponding $\tau(\mathrm{q})$. As a result, $\tau(\mathrm{q})$ versus $q$ is derived as shown in Fig. 10d. Note that red dots represents the values derived with the use of the WTMM method, and blue lines are theoretical curves $\tau(\mathrm{q}) = -\log_4(|\mathrm{p}_1|^{\mathrm{q}} + |\mathrm{p}_2|^{\mathrm{q}} + |\mathrm{p}_3|^{\mathrm{q}} + |\mathrm{p}_4|^{\mathrm{q}}) = -\log_4(0.69^{\mathrm{q}} + 0.46^{\mathrm{q}} + 0.46^{\mathrm{q}} + 0.31^{\mathrm{q}})$. Fig. 10d shows that the spectrum $\tau(q)$ from the WTMM method matches with the theoretical curves. Furthermore, the multifractal spectrum $\mathrm{D(h)}$ is derived from $\tau(\mathrm{q})$ through a Legendre transform. Fig. 10e shows the multifractal spectrum $\mathrm{D(h)}$ vs. the singularity exponent h. By comparing the WTMM results with the theoretical curves, it is evident that the WTMM method is effective to extract multifractal spectrums from a nonlinear time series.

### 3.2 Recurrence quantification analysis

Recurrence (i.e., approximate repetitions of a certain event) is one of the most common phenomena in natural and engineering systems. For examples, the human heart is near-periodically beating to maintain vital living organs, and manufacturing machines are cyclically forming sheet metals during production. Real-time sensing brings the proliferation of big data (i.e., dynamic, nonlinear, nonstationary, high dimensional) from complex processes. This provides an unprecedented opportunity for data-driven characterization and modeling of nonlinear recurrence behaviors towards system informatics and control. However, most of existing approaches adopt linear methodologies for analyzing dynamic recurrences. Traditional linear methods interpret the regular structure, e.g., dominant frequencies in the signals. They have encountered certain difficulties to capture the nonlinearity, nonstationarity and high-order variations. For example, Fourier analysis does not provide the temporal localization of frequency components, and assume spectral components exist at all times (i.e., stationarity). Instead, system diagnostics and process control are more concerned with aperiodic recurrences and nonlinear recurrence variations.

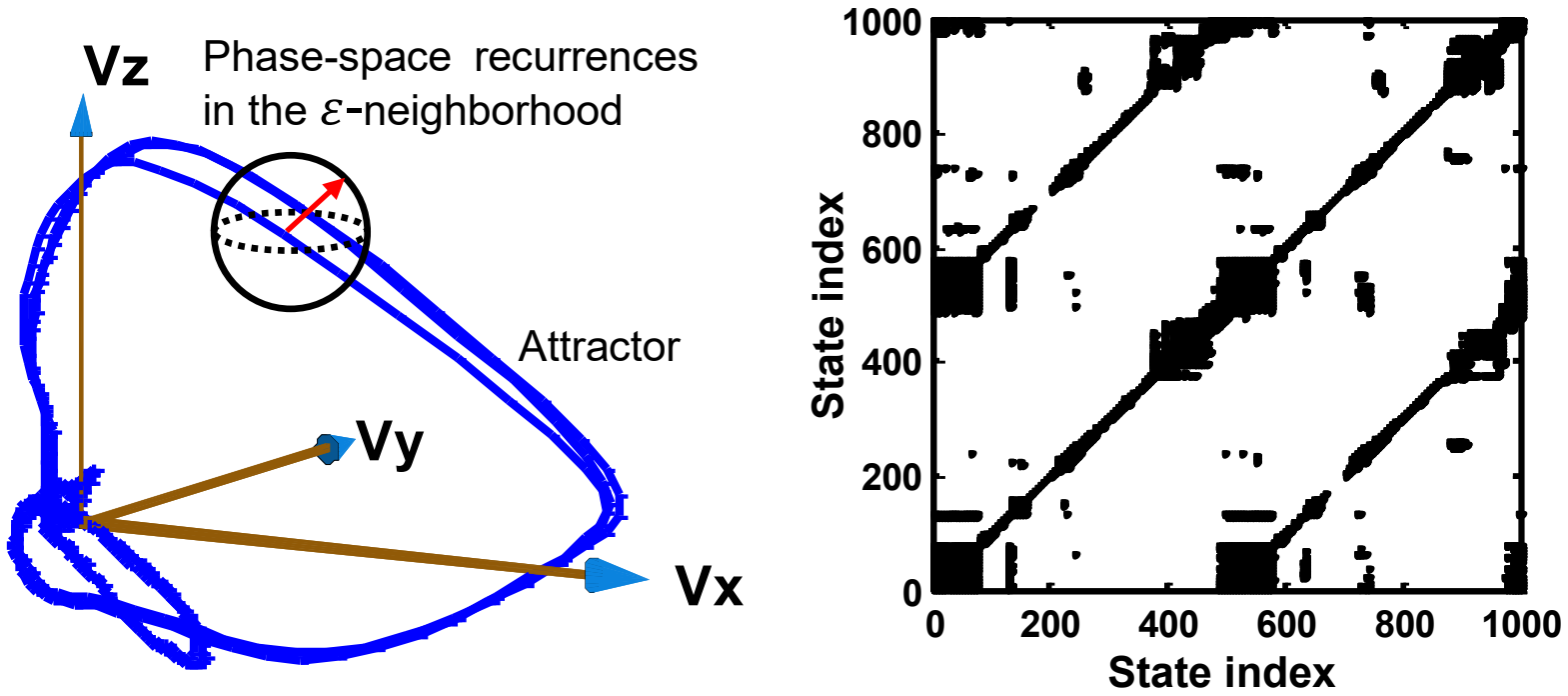

**Fig. 11:** (a) An example of ECG trajectories in the 3-dimensional phase space; (b) The recurrence plot characterizes the proximity of two states $\vec{x}(i)$ and $\vec{x}(j)$, i.e., $R(i,j) \coloneqq \Theta(\varepsilon - \|\vec{x}(i) - \vec{x}(j)\|)$, where $\Theta$ is the Heaviside function and $\|\cdot\|$ is a distance measure.

Therefore, nonlinear recurrence methodologies are urgently needed to handle the underlying complexity in the big data. Poincaré recurrence theorem shows that if a dynamical system has the measure preserving transformation, its trajectories eventually reappear in the $\varepsilon$-neighborhood of former states [4]. The methodology of nonlinear recurrence analysis is emerged from the theory of nonlinear dynamics, and characterizes recurrence behaviors in the high-dimensional state space. The recurrence plot was introduced by Eckmann *et al.* in the late 1980's [23] to characterize the proximity of states in the phase space. Mathematically, the recurrence plot is defined as: $\boldsymbol{R}(i,j) = \Theta(\varepsilon - \|\vec{\boldsymbol{x}}(i) - \vec{\boldsymbol{x}}(j)\|)$, where $\Theta$ is the Heaviside function, $\varepsilon$ the neighborhood size and $||\cdot||$ is a distance measure. For the lag-reconstructed phase space $\vec{\boldsymbol{x}}(i) = \left(x_i, x_{i+\tau}, \cdots, x_{i+\tau(M-1)}\right), i = 1, \cdots, N - \tau(M-1)$, which is lag-reconstructed from a time series, the computation of recurrence plot will be

$$\boldsymbol{R}(i,j) = \Theta\left[\varepsilon - \sqrt{\sum_{m=0}^{M-1}\left(x_{i+m\tau} - x_{j+m\tau}\right)^2}\right]$$

It is worth mentioning that if the neighborhood size $\varepsilon$ is too small, there will be few recurrence points in the plot. Thus, we can hardly learn anything about recurrence structures. However, if $\varepsilon$ is too large, almost every point is a neighbor of every other point. In the literature, there are several "rules of thumb" for the selection of $\varepsilon$: (1) a small percentage of the maximum diameter of the state space; (2) a fixed scale region in the recurrence rate; (3) fix the number of neighbors for every point; (4) take into account the standard deviation of the observational noise. An optimal choice of $\varepsilon$ facilitates the characterization of recurrence structures and dynamical properties of complex systems.

As shown in Fig. 11, recurrence plot captures topological relationships in the state space as a 2D image. If two states are located close to each other in the $m$-dimensional state space (e.g., 3D space in Fig. 11a), the color code is black (Fig. 11b). If they are located farther apart, the color is white. The structure of a recurrence plot has distinct topology and texture patterns (Fig. 11b). The ridges locate the nonstationarity and/or the switching between local behaviors. The parallel diagonal lines indicate the near-periodicity of system behaviors. Recurrence quantification analysis (RQA) measures intriguing structures and patterns in the recurrence plot, including small-structures (e.g., small dots, vertical and diagonal lines), chaos-order transitions, as well as chaos-chaos transitions. A comprehensive review on recurrence quantifiers is reported in Marwan et al., 2007 [24]. Here, we will present several examples of recurrence quantifiers as follows: (1) Recurrence rate ($RR$) - a measure of the density of recurrence points in the RP, $\mathrm{RR} = \frac{1}{N^2}\sum_{i,j=1}^{N} R(i,j)$, where N is the number of states in the attractor; (2) determinism ($DET$) - the percentage of recurrence points which form the diagonal lines, $\mathrm{DET} = \frac{\sum_{l=l_{min}}^{N} lP(l)}{\sum_{l=1}^{N} lP(l)}$, where $P(l)$ is the histogram of diagonal line length; (3) entropy ($ENT$) - Shannon information entropy for the probability distribution of the diagonal line lengths $P(l)$, $ENT = -\sum_{l=l_{min}}^{N} p(l)\ln p(l)$; (4) laminarity ($LAM$) - the percentage of recurrence points which form vertical lines, $LAM = \frac{\sum_{v=v_{min}}^{N} vP(v)}{\sum_{v=1}^{N} vP(v)}$, where $P(v)$ is the histogram of vertical line lengths; (5) trapping time (TT) - the average length of vertical structures, $\mathrm{TT} = \frac{\sum_{v=v_{min}}^{N} vP(v)}{\sum_{v=v_{min}}^{N} P(v)}$. Recurrence quantification analysis goes beyond the visual inspection in the RP and provides complexity measures of system dynamics. If we compute the RQA measures in small windows (sub-matrices) along the line of identity (LOI) of the RP, time-dependent behaviors of system dynamics will be quantified. Recurrence quantification analysis have successful applications in various disciplines, e.g., physiology [25-28], biology [29], economy [30], manufacturing [31], geophysics [32], and neuroscience [28, 33].

### 3.3 Multiscale recurrence quantification analysis

However, complex systems exhibit recurrence characteristics in multiple spatial and temporal scales. Most of existing recurrence methods considered the data in a single scale. In addition, recurrence computation is highly expensive, i.e., a squared increase (i.e., $O(n(n-1)/2)$) with the size of data $n$ [3]. This limits the use of recurrence approaches for big data, which are *often collected in real-time monitoring of complex systems*. Multiscale analysis is of fundamental importance to solve engineering and physics problems that have important characteristics in spatial, temporal and/or frequency scales. For examples, multiscale spatial models were developed for the prediction of weather evolution in meteorology [34] and the control of nanomaterial growth in material science [35]. Hilbert-Huang transform empirically decomposes time series into intrinsic mode functions (IMFs) via the sifting process, thereby captures multiple instantaneous frequencies in the data [36, 37]. In addition, wavelet transform elucidates multiscale information of sensing data in the time-frequency domain [17, 38].

However, very little work has been done to investigate multiscale recurrence dynamics. Our previous research developed a novel multiscale framework to quantify recurrence dynamics in complex systems and resolve computational issues for large-scale datasets [3, 39]. As opposed to traditional single-scale recurrence analysis, we characterize and quantify recurrence dynamics in multiple wavelet scales. As shown in Fig. 12, wavelet transform decomposes nonstationary VCG signals into various frequency bands for effectively separating the system transient, intermittent and steady behaviors. Wavelet packet decomposition (WPD) introduces both the wavelet function and scaling function for an efficient pyramid decomposition of signal space $V_j$ into an approximation space $V_{j+1}$ and a detail space $W_{j+1}$. The approximation space $V_{j+1}$ and the detail space $W_{j+1}$ are, then, divided iteratively in the next level. This provides a better resolution in both time and frequency scales. As shown in Fig. 12, the time series $\boldsymbol{X} = \{x_1, x_2, \cdots, x_N\}^T$, denoted as $\boldsymbol{W}_{0,0}$, is passed through the lowpass filter $G(\cdot)$ and highpass filter $H(\cdot)$ and followed by the dyadic subsampling process in each level. This subband coding is repeated to produce the $k$th level coefficient sets that are denoted as $\boldsymbol{W}_{k,n}, n = 0, \cdots, 2^k - 1$. The redundancy is removed because each set $\boldsymbol{W}_{k,n}$ is of length $N/2^k$ and the total length in level $k$ is the same as the original time series $\boldsymbol{X}$. Note that a narrower wavelet function captures high-frequency transient behaviors in a fine-grained time resolution, and the wider one characterizes low-frequency steady behaviors in a better frequency resolution. Thus, a nonstationary time series is resolved into multiple non-overlapping frequency bands. In addition, the dyadic subsampling in discrete wavelet transforms leads to the reduction of sample size. Notably, wavelet decomposition and subsampling do not loss any information. The original long-term signal can be perfectly reconstructed from wavelet coefficients.

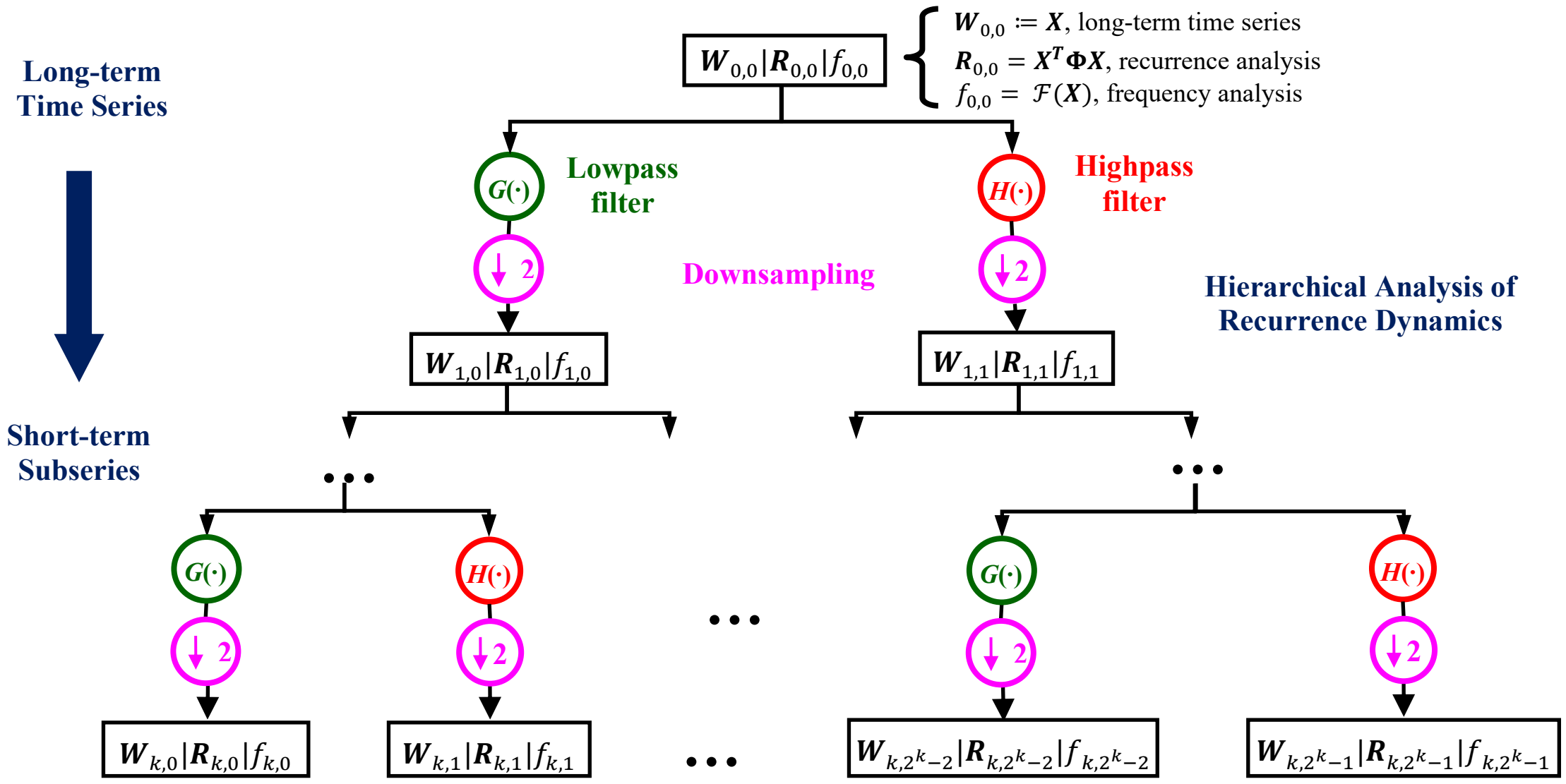

**Fig. 12:** Flow diagram illustrating the WPD of a long-term time series $\boldsymbol{X}$, as well as the hierarchical analysis of recurrence and frequency behaviors. The subband coding (i.e., lowpass filter $G(\cdot)$ and highpass filter $H(\cdot)$) and dyadic subsampling processes decompose a long-term time series $\boldsymbol{X}$, defined as $\boldsymbol{W}_{0,0}$, into the $k^{th}$ level subseries that are denoted as $\boldsymbol{W}_{k,n}, n = 0, \cdots, 2^k - 1$, thereby facilitating the analysis of recurrence dynamics [40].

Multiscale recurrence analysis integrates recurrence quantification analysis into the framework of wavelet subband coding. In each wavelet scale, recurrence analysis further quantifies nonlinear system dynamics. Our previous research showed that the recurrence plot $R_{0,0}(i,j)|_{i,j=1,\cdots,N}$ of original time series $\boldsymbol{X}$ can be perfectly reconstructed with the $k^{th}$ level of wavelet coefficients $\boldsymbol{W}_{k,n}$ and their recurrence plots $R_{k,n}(i,j)$ [3]. Dynamical properties in the original recurrence plot are preserved after the wavelet packet decomposition. Recurrence dynamics pertinent to the original time series are further delineated by computing the recurrence plots from the wavelet subseries $\boldsymbol{W}_{k,0} \cdots \boldsymbol{W}_{k,2^k-1}$. Note that many previous approaches adjusted the threshold $\varepsilon$ for an optimal recurrence plot in the presence of noises. Multiscale recurrence analysis is more robust to observational noises because it decomposes the system behaviors into different frequency bands. For example, noises will be separated into the high-frequency band and long-term trend will go into the low-frequency band. When there is a mixture of noise, nonlinear and nonstationary behaviors, multiscale wavelet decomposition separates the mixture of information into various wavelet scales. This further reduces the complexity of nonlinear dynamics within each scale. These shorter wavelet subseries make expensive recurrence computations not only plausible but also more effective within wavelet scales. Multiscale recurrence analysis facilitates the prominence of hidden recurrence properties that are usually buried in a single scale.

## 4. Healthcare applications

Human heart is essentially an autonomous electro-mechanical blood pump that operates near-periodically to maintain vital living organs. ECG signals contain a wealth of dynamic information pertinent to cardiac operations, which is indispensable for cardiac care—from monitoring and diagnosis to treatment planning to smart health management. One lead ECG captures 1-dimensional temporal view of space-time cardiac electrical activity. Multi-lead ECG systems provide multi-directional views of such space-time dynamics [2]. A normal ECG tracing is often segmented into P wave, QRS complex, and T wave (see Fig. 1b). Atrial depolarization (and systole) is represented by the P wave, ventricular depolarization (and systole) is represented by the QRS complex, and ventricular repolarization (and diastole) is represented by the T wave [41]. In addition, heart rate variability (HRV) refers to the fluctuations in the sequential heart-beat intervals, also called RR intervals. Heart-beat dynamics are highly pertinent to the function of autonomic nervous system. Autonomic nervous control brings a greater level of nonlinear dynamics in the presence of nonstationarity and noises. Most existing work focused on the analysis of time-domain ECG signals from a single sensor. Time-domain algorithms were usually developed to quantify the characteristics of ECG wave deflections (i.e., P, QRS and T waves) [42-44]. Examples of ECG features include PR interval, RR interval, ST elevation/depression, QT interval, R amplitude. Also, Fourier analysis was utilized to transform time-domain ECGs to extract hidden features in the frequency domain [45-47]. However, Fourier analysis does not provide temporal location of frequency components, and assumes spectral components exist at all times (i.e., stationarity). Nonstationarity in cardiovascular systems fueled increasing interests in wavelet analysis of ECG signals to delineate local time and frequency information for applications such as adaptive representation [29], PQRST segmentation [48-50], noise cancellation [38], and arrhythmia recognition [51]. Further, nonlinear methods were developed to reconstruct the phase space from 1-lead ECG and then characterize the dynamics of cardiovascular systems [26, 52].

However, many previous works underuse multi-lead ECG signals and overlook spatio-temporal dynamics in the heart. Multiple sensors at various locations on the human body respond to process changes differently. Time-domain ECG—a projected view of space-time cardiac electrical activity—diminishes important spatial information pertinent to tissue damages in the heart (e.g., myocardial infarction). Most existing methods are influenced by such an information loss, thereby failing to extract effective ECG biomarkers sensitive to cardiac malfunctions. The objective of this study reported in this section is to present two case studies on the characterization and modeling of nonlinear dynamics in cardiovascular systems. First, the approach of wavelet multifractal analysis is developed to quantify nonlinear dynamics in heart rate time series. These fractal features provide useful information about nonlinear scaling behaviors and the complexity of autonomic cardiovascular function. Second, we will present a novel multiscale recurrence approach to study disease-altered nonlinear dynamics in the spatiotemporal vectorcardiogram (VCG) signals. As opposed to the traditional single-scale recurrence analysis, we characterize and quantify recurrence behaviors within multiple wavelet scales. Also, wavelet dyadic subsampling makes the expensive recurrence computations not only plausible for the long-term time series but also more effective under the stationary assumptions in multiple wavelet scales.

### 4.1 Nonlinear characterization of heart rate variability

Heart rate variability (HRV) analysis plays an important role in the detection of disorders in autonomic cardiovascular function. Since the 1980s, linear and frequency-domain approaches are widely used in the HRV analysis but are limited in the ability to capture nonlinear dynamics in the long-term HRV time series. For example, Fourier analysis is efficient to transform data from time domain to frequency domain but does not provide the temporal localization of frequency components. Also, linear statistical methods, e.g., analysis of variance (ANOVA), have certain difficulties to capture the nonlinearity, nonstationarity and high-order variations. Therefore, linear methods tend to bring less realistic characterization and quantification of nonlinear time series. Notably, recent research showed that congestive heart failure, a major life-threatening cardiac disorder, leads to a loss of multifractality [16]. Heart failure is caused by a loss of cardiac ability to supply sufficient blood flows to the body. As a result, central nervous system controls the heart rate to compensate heart failure by maintaining blood pressure and perfusion, e.g., increasing the sympathetic activity. However, autonomic cardiovascular function is not only nonlinear and nonstationary but also with long-range correlations, at time scales ranging from seconds to minutes to hours. This is significantly different from acute cardiac events pertinent to only a segment of ECG signals. Therefore, long-term time series are necessary to delineate the complex long-range dependence behaviors in multiple scales for the identification of heart failures. Fig. 13 shows examples of scaling exponents function and multifractal spectrum extracted from HRV time series of healthy control and heart failure subjects. Scaling exponents $\tau(q)$ of the healthy subject (blue dots) are more linear than those of heart failures (red crosses). Multifractal spectrum $D(h)$ is obtained through a Legendre transform from the $\tau(q)$ in Fig. 13a. It is worth mentioning that multifractal spectrum $D(h)$ for the heart failure subject is narrower than healthy control, indicating the loss of multifractality.

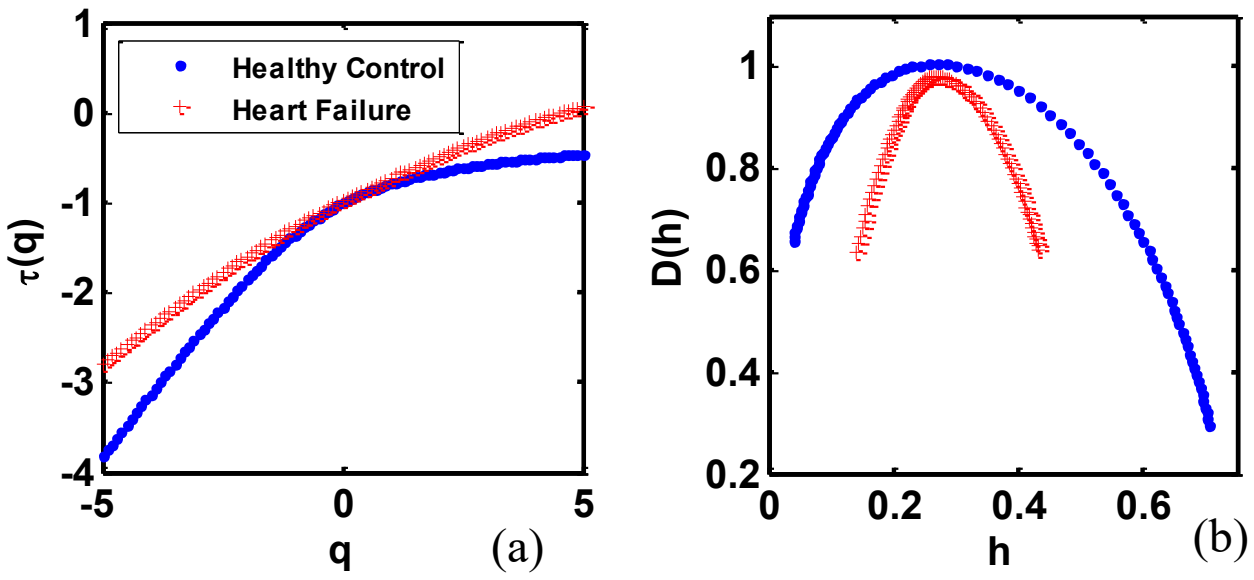

**Fig.13**: (a) scaling exponents function: $\tau(q)$ versus $q$ and (b) multifractal spectrum: $D(h)$ versus $h$ extracted from heart rate variability time series of healthy control and heart failure subjects.

This section presents our previous efforts on characterization and modeling of nonlinear dynamics in HRV time series and further evaluating their classification performances. For that purpose, we used three well-known classification algorithms, namely logistic regression, k-nearest neighbor and artificial neural network. We build three classification models for nonlinear features to benchmark the performance in detecting disorders of autonomic cardiovascular function. In this study, we analyzed the 24-hour heart rate time series that are gathered from 54 healthy control (HC) subjects and 29 congestive heart failure (CHF) patients, available in the PhysioNet [53]. Heart rate time series is preprocessed to eliminate erroneously large intervals and outliers due to missed beat detections following the same procedure as in [16]. The preprocessing procedures include (a) a moving-window average filter, and (b) increment smoothing. For the 5 consecutive points in a moving window, the central point is removed if it is greater than twice the local mean calculated from the other four points. There is no interpolation in this moving-window average filter. The second step calculates differences between adjacent elements in the time series. If the successive increments has opposite sign with amplitudes > 3×standard deviation of increment series, both increments will be replaced by the interpolated value in between. The new heart rate time series is, then, reconstructed from the post-processed series of increments.

**Feature extraction:** We have utilized two alternative approaches, namely wavelet multifractal analysis and multiscale recurrence analysis, to extract nonlinear dynamic features from long-term heart rate time series. Features extracted in the wavelet multifractal analysis include multifractal spectrum $\tau(q)$ and fractal dimension $D(h)$. The fractal features provide useful information about nonlinear scaling behaviors and the complexity of autonomic cardiac function. For multiscale recurrence analysis, six recurrence statistics, namely RR, DET, LMAX, ENT, LAM and TT are exacted to quantify the nonlinear recurrence behaviors in wavelet subseries. Therefore, a total of $6 \times 2^k$ recurrence features are exacted for the $k^{\text{th}}$ level wavelet packet decomposition. In the case study of HRV, $k$ is chosen from 6 to 9 for all subjects to explore the optimal decomposition level that captures the frequency ranges of disease variations. The neighborhood size $\varepsilon$ in recurrence plots was chosen to be 5% of the maximal distance of state space. The total length of each HRV recording is pruned to be 76000 data points to keep computational consistency for all subjects.

**Feature selection:** The method of sequential feature selection is used to optimally choose a subset of features that are closely correlated with the disease variations [3, 20]. Note that a large amount of features are extracted from three nonlinear approaches. As a result, this may bring the "curse of dimensionality" issues for classification models, e.g., increased model parameters and overfitting problems [3, 40, 54]. In addition, such a high dimensional feature space hinders the development of a deeper understanding of cardiac pathology. Hence, we use the strategy of sequential forward feature selection to optimally choose a subset of features that are strongly correlated with process variations. Starting from an empty feature subset, an additional feature $\mathscr{s}^+$ is selected when it maximizes the objective function $\mathcal{J}(\mathcal{S}_\ell + \mathscr{s}^+)$, which wraps the classification model. This process is repeated until it reaches the desired subset size. Feature selection not only surmounts the model complexity and overfitting problems, but also provides faster and more cost-effective models with the optimal feature subset.

Table 1: Unpaired $t$-test and KS test for selected features

| Statistic Tests | Analysis Methods* | Test Statistics | | | | | | | | | |
|---|---|---|---|---|---|---|---|---|---|---|---|
| | | 1st | 2nd | 3rd | 4th | 5th | 6th | 7th | 8th | 9th | 10th |
| Unpaired $t$-test ($p$-value) | WMA | 6.2e-04 | 3.5e-03 | 0.012 | 8.9e-04 | 3.1e-03 | 0.030 | 8.5e-03 | 0.037 | 0.020 | 3.5e-03 |
| | MRA | 2.5e-03 | 4.9e-03 | 1.8e-03 | 8.3e-04 | 0.017 | 2.5e-05 | 5.7e-03 | 0.251 | 0.153 | 0.050 |
| Two-sample KS test (KS statistic) | WMA | 0.613 | 0.372 | 0.324 | 0.467 | 0.425 | 0.343 | 0.430 | 0.343 | 0.375 | 0.433 |
| | MRA | 0.452 | 0.391 | 0.340 | 0.427 | 0.396 | 0.429 | 0.385 | 0.346 | 0.335 | 0.305 |

*WMA – Wavelet Multifractal Analysis; MRA: Multisccale Recurrence Analysis

**Feature analysis:** As shown in Table I, we evaluated the individual feature separately using two statistical tests, namely unpaired $t$-test and Kolmogorov-Smirnov (KS) test. There are 10 features for each method that are optimally chosen by the feature selection algorithms. In the unpaired $t$-test, the smaller $p$-value indicates more evidences to reject the null hypothesis, i.e., the feature has the same distribution between HC and CHF groups. In the KS test, a larger KS statistic shows that this feature has more distinct cumulative distribution functions between the HC and CHF groups. Table I shows that two statistic tests agree on the fact that most of the features are significant, because the majority of $p$-values are $<0.05$ and KS statistic $>0.3$. However, 1-dimensional statistical test does not account for the feature dependence in the high-dimensional space.

**Classification performance:** Therefore, we carried out classification experiments with two groups of features to evaluate the combinatorial effects of multi-dimensional features. Three classification models are K-Nearest-Neighbor (KNN), logistic regression (LR) and artificial neural network (ANN). As shown in Fig. 14, the bar plot is used to visualize the statistics of classification performance (i.e., sensitivity, specificity and accuracy) that are computed from 100 random replications of the 4-fold cross-validation.

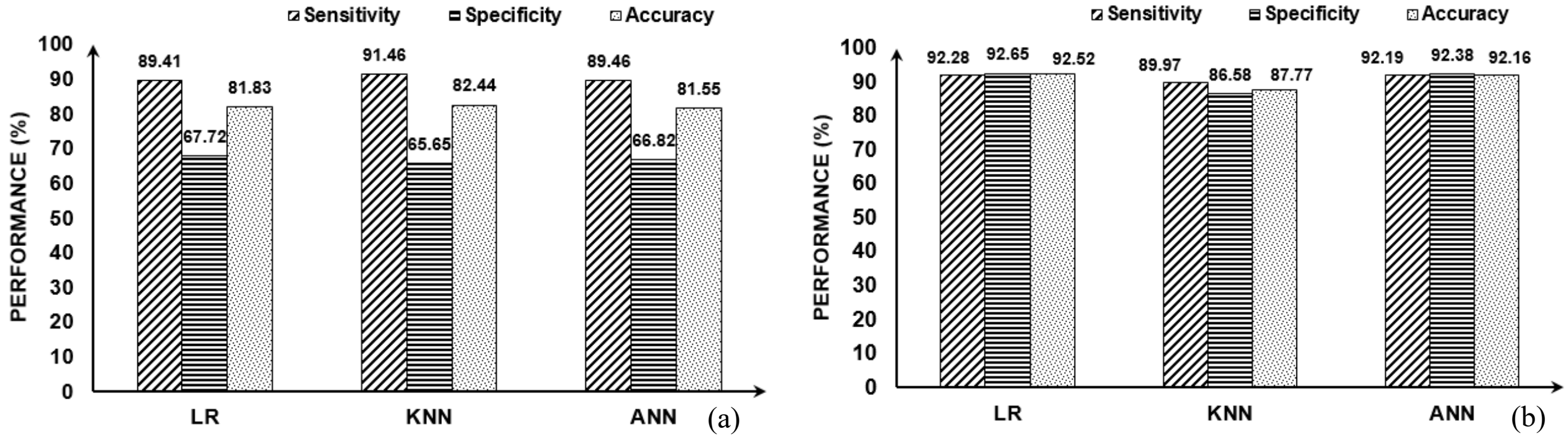

**Fig.14**: Performance results for (a) wavelet multifractal features and (b) multiscale recurrence features using three classification models – logistic regression (LR), K-nearest neighbors (KNN) and artificial neural network (ANN).

Fig. 14a shows the sensitivity, specificity and accuracy respectively for features extracted from wavelet multifractal analysis of heart rate time series. Fig. 14a demonstrates an average sensitivity of 89.41%, a specificity of 67.72%, and an accuracy of 81.83% for the logistic regression. In addition, KNN and ANN models yielded approximately similar results but with small deviations. Overall, the KNN model was shown to achieve a better accuracy (i.e., 82.44%) than the other two models. Experimental results of three classification models show that the features extracted from wavelet multifractal analysis are significant between CHF and HC subjects.

Fig. 14b present the classification results for features extracted from multiscale recurrence analysis of heart rate time series. Notably, multiscale recurrence features lead to generally better results for all three classification models than features extracted from wavelet multifractal analysis. Logistic regression models were shown to further improve the sensitivity to 92.28%, the specificity to 92.65%, and the accuracy to 92.52%. The ANN models are shown to yield approximately the same results (i.e., an accuracy of 92.16%) as logistic regression, but the performance of KNN models is lower than both ANN and logistic regression models. Overall, multiscale recurrence analysis delineates nonlinear and nonstationary behaviors in multiple scales of heart rate time series, and is shown to yield better results for the classification of healthy control and heart failure subjects.

### 4.2 Multiscale recurrence analysis of space-time physiological signals

The human heart is a 3-dimensional object and cardiac electrical activities are near-periodically conducting across space and time. The electrocardiogram (ECG) contains a wealth of dynamic information pertinent to cardiac functioning, but 1-lead ECG only captures one directional view of spatiotemporal heart activities. In contrast, 3-lead vectorcardiogram (VCG) monitors the spatiotemporal cardiac electrical activity along three orthogonal X, Y, Z planes of the body, namely, frontal, transverse, and sagittal [39]. However, 3-lead VCG is not as commonly used as 12-lead ECG because medical doctors are accustomed to using the time-domain ECG in clinical applications. Dower *et al.* [55, 56] and our previous study [57] showed that 3-lead VCG can be linearly transformed to 12-lead ECG without a significant loss of clinically useful information. Thus, 3-lead VCG surmounts not only the information loss in 1-lead ECG but also the redundant information in 12-lead ECG.

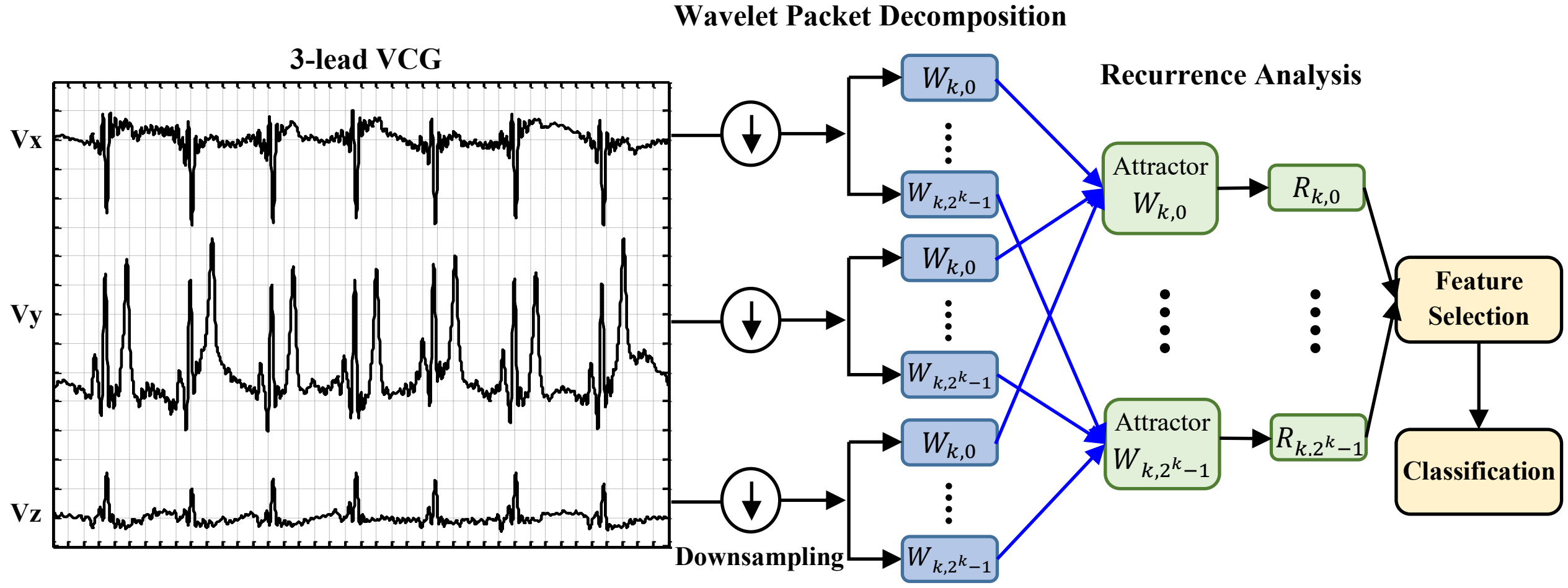

**Fig. 15:** Multiscale recurrence analysis of disease-altered VCG signals.

However, most of previous nonlinear methods only considered the lag-reconstructed state space from 1-lead ECG signals. Although 3-lead VCG provides a new way to investigate the cardiac dynamical behaviors, few previous approaches have studied the disease-altered recurrence dynamics in the space-time VCG signals. This present paper developed a novel multiscale recurrence approach to not only explore recurrence dynamics but also resolve the computational issues for the large-scale datasets. As shown in Fig. 15, the long-term VCG signal, followed by the dyadic subsampling, is decomposed into wavelet subseries. Each subseries is iteratively decomposed to produce $2^k$ subsets of wavelet sub-signals, denoted as $\boldsymbol{W}_{k,n}, n = 0, \cdots, 2^k - 1$, in the $k^{th}$ level. Within each wavelet scale, recurrence analysis is utilized to quantify the underlying dynamics of nonlinear systems. We performed multiscale recurrence analysis of 448 VCG recordings (368 MIs and 80 HCs) available in the PhysioNet PTB Database [53]. Each recording contains 15 simultaneous heart-monitoring signals, i.e., the conventional 12-lead ECG and the 3-lead VCG.

Notably, we have previously extracted RQA features from the 3-lead VCG in the original scale for the identification of myocardial infarction subjects [58]. In addition, we utilized the DWT to decompose VCG signals into multiple wavelet scales, and compute RQA features from not only the original single scale but also multiple wavelet scales [39]. It is worth mentioning that only 4000 data points in the 3-lead VCG are used for the single-scale and DWT recurrence analysis due to the computational complexity. In the study reported in this section, we further utilized wavelet packets decomposition for not only quantify multiscale recurrence dynamics but also resolve the computational issues for large-scale datasets. It may be noted that the 3-lead VCG of 16000 data points are utilized for recurrence quantification analysis in this present study with the use of WPD dyadic sampling.

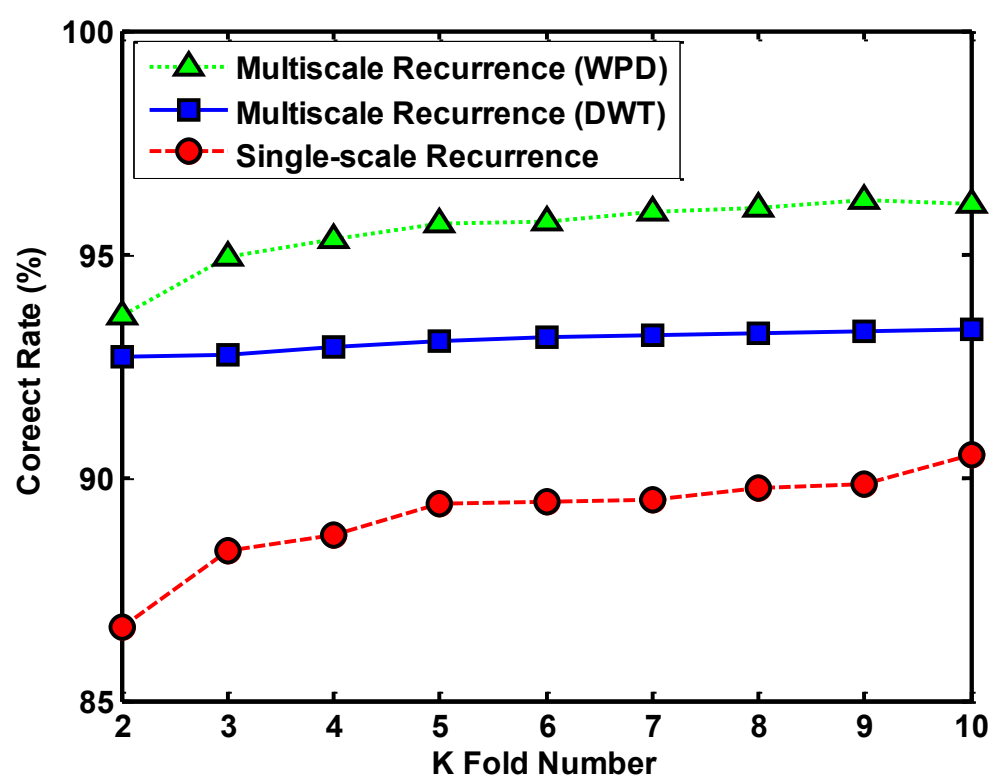

**Fig. 16**: The comparison of classification performance between single-scale and multi-scale recurrence analysis.

As shown in Fig. 16, multiscale recurrence analysis (i.e., DWT and WPD) show better performances (in terms of correct rates) than the single-scale recurrence analysis. The correct rate using DWT recurrence analysis (93.2% from 10-fold cross validation) is 2.7% higher than the single-scale recurrence analysis (90.5% from 10-fold cross validation). Moreover, the proposed WPD recurrence analysis increases the correct rate about 2.9% from the previous DWT recurrence analysis. The correct rate for the identification of MI subjects is 96.1% in the WPD recurrence analysis, which is about 5.6% increase from the single-scale analysis.

In the literature, little has been done to investigate multi-scale variations of phase-space recurrences underlying the space-time VCG signals. Previous work focused primarily on the recurrences in time-domain signals from a single sensor. In addition, existing recurrence methods only considered the data in a single scale. Our previous research developed a novel multi-scale framework to characterize and quantify the dynamics of transient, intermittent and steady recurrences within wavelet scales. Multiscale recurrence analysis facilitates the prominence of hidden recurrence properties that are usually buried in a single scale. Our previous research results are summarized as follows: **(1) *Single-scale vs. multi-scale recurrence analysis:*** As opposed to the traditional recurrence analysis in a single scale, we delineate the recurrence dynamics into multiple wavelet scales. **(2) *Long-term recurrence analysis:*** Few, if any, previous approaches have been capable of quantifying the recurrence dynamics from a long-term time series. Recurrence computation is highly expensive (i.e., $O(n(n-1)/2)$) as the size of time series $n$ increases. The dyadic subsampling in wavelet packet decomposition effectively resolves the computational issues for the large-scale recurrence analysis. **(3) *Disease-altered recurrence dynamics*:** It is shown that recurrence dynamics are significantly different in wavelet scales between healthy control (HC) and myocardial infarction (MI) subjects. Multiscale recurrence analysis identifies the MI with an average sensitivity of 96.8% and specificity of 92.8%, which is much better (i.e., 5.6% increase) than the single-scale recurrence analysis.

## 5. Summary

Real-world physiological systems show high level of nonlinear and nonstationary behaviors in the presence of extraneous noises. Nonlinear dynamic methods provide significant opportunities to explore the hidden patterns and relationships in complex physiological systems. This chapter presents a review of theoretical developments and tools of nonlinear dynamics principles as well as their applications in healthcare data analytics. Specifically, we showed the methodological details of multifractal spectrum analysis and multiscale recurrence analysis with case studies in modeling and analysis of heart rate variability and space-time ECG signals. From the foregoing, it is evident that healthcare data analytics can be greatly advanced from using sensor-based characterization and modeling of nonlinear dynamics.

We first introduced the methodology of multifractal spectrum analysis and its applications to identify congestive heart failure subjects using the 24-hour heart rate time series. Experimental results demonstrated the effectiveness to delineate nonlinear and nonstationary behaviors in multiple scales of time series. For the multifractal features, the logistic regression models achieve a sensitivity around 89.41% and an average specificity of 67.72%. KNN and ANN models yielded approximately similar results but with small deviations. The multifractal approach was shown to effectively capture nonlinear dynamic behaviors in the 24-hour heart rate time series. In addition, logistic regression models were shown to further improve the sensitivity to 92.28%, the specificity to 92.65%, and the accuracy to 92.52% using features extracted from multiscale recurrence analysis of heart rate time series. In general, Overall, multiscale recurrence analysis delineates nonlinear and nonstationary behaviors in multiple scales of heart rate time series, and is shown to yield better results for the classification of healthy control and heart failure subjects than wavelet multifractal methods.

Furthermore, this chapter presents a novel multiscale recurrence approach to analyze the 3-lead VCG signals for the detection of MIs. Few, if any, previous work studied disease-altered nonlinear dynamics hidden in long-term spatiotemporal VCG signals. Notably, most of existing nonlinear dynamic methods considered the time-delay reconstructed phase space from 1-dimensional time series for the investigation of physiological dynamics. Computer experiments demonstrate that the proposed approach yields better performances by characterizing the nonlinear and nonstationary behaviors in multiple wavelet scales. Multiscale recurrence analysis of VCG signals leads to a superior classification model that detects the myocardial infarction with an average sensitivity of 96.8% and specificity of 92.8%, which is much better (i.e., 5.6% increase in terms of correct rates) than the single-scale recurrence analysis.

The theory of nonlinear dynamics has been primarily studied in mathematics and physics. Most of previous works have begun the adaptation of the existing results in nonlinear dynamics body of knowledge into healthcare data analytics. However, realizing the full potential of nonlinear dynamics theory for healthcare analytics calls upon the new advancement of nonlinear dynamics methodologies, as well as integration of existing nonlinear methods with healthcare analysis tools. For example, very little has been done to adapt nonlinear dynamics principles into operational analytics in healthcare systems engineering. Also, nonlinear dynamics researchers have traditionally not addressed the issues of how to construct nonlinear models from the wealth of process data, and how to address noises in real-world healthcare processes. These research problems are critically important to improving the performance of healthcare systems and achieving a remarkable reduction of healthcare costs. Future research efforts addressing these problems will advance not only current healthcare practice, but also will enrich the theory of nonlinear dynamics and further expand its research domain to health care. We hope that our limited and focused review will inform subsequent studies that will focus on the development of novel nonlinear dynamics methodologies for improving healthcare services and optimizing healthcare systems that are so vitally important for smart health.

## 6. Acknowledgments

The authors thank the National Science Foundation (CMMI-1266331, IIP-1447289 and IOS-1146882) for support the research presented in this book chapter. In addition, the author (Hui Yang) acknowledges the support of his PhD advisors, Dr. Satish T.S. Bukkapatnam and Dr. Ranga Komanduri, and his collaborator, Dr. Eric S. Bennett without whose efforts this work would not have been possible.

## 7. References

[1] American Heart Association Writing Group, "Heart Disease and Stroke Statistics—2014 Update: A Report From the American Heart Association," *Circulation,* vol. 129, pp. e28-e292, 2014.

[2] H. Yang, S. T. S. Bukkapatnam and R. Komanduri, "Spatio-temporal representation of cardiac vectorcardiogram signals," *BioMedical Engineering Online,* vol. 11, 2012.

[3] Y. Chen and H. Yang, "Multiscale recurrence analysis of long-term nonlinear and nonstationary time series," *Chaos, Solitons & Fractals,* vol. 45, pp. 978-987, 2012.

[4] A. Katok and B. Hasselblatt, *Introduction to the Modern Theory of Dynamical Systems.* Cambridge University Press, 1995.

[5] Cardiac Arrhythmia Suppression Trial Investigators, "Preliminary Report: Effect of Encainide and Flecainide on Mortality in a Randomized Trial of Arrhythmia Suppression after Myocardial Infarction," *N. Engl. J. Med.,* vol. 321, pp. 406-12, 1989.

[6] J. Q. You and F. Nori, "Atomic physics and quantum optics using superconducting circuits," *Nature,* vol. 474, pp. 589-97, 2011.

[7] J. R. Pratt and A. H. Nayfeh, "Design and Modeling for Chatter Control," *Nonlinear Dynamics,* vol. 19, pp. 49-69, 1999.

[8] R. Roy, T. W. Murphy, T. D. Maier, Z. Gills and E. R. Hunt, "Dynamical control of a chaotic laser: Experimental stabilization of a globally coupled system," *Phys. Rev. Lett.,* vol. 68, pp. 1259-62, 1992.

[9] M. Ishikawa, "Precise fabrication of nanomaterials: A nonlinear dynamics approach," *Chaos: An Interdisciplinary Journal of Nonlinear Science,* vol. 15, pp. 047503, 2005.

[10] R. Tenny, L. S. Tsimring, L. Larson and H. D. I. Abarbanel, "Using Distributed Nonlinear Dynamics for Public Key Encryption," *Phys. Rev. Lett.,* vol. 90, pp. 047903, 2003.

[11] F. Takens, "Detecting strange attractors in turbulence," in *Dynamical Systems and Turbulence, Warwick 1980, Lecture Notes in Mathematics,* Anonymous Springer-Verlag, 1981.

[12] M. B. Kennel, R. Brown and H. D. I. Abarbanel, "Determining embedding dimension for phase-space reconstruction using a geometrical construction," *Phys. Rev. A,* vol. 45, pp. 3403-3411, 1992.

[13] A. M. Fraser and H. L. Swinney, "Independent coordinates for strange attractors from mutual information," *Phys. Rev. A,* vol. 33, pp. 1134-1140, February, 1986.

[14] B. B. Mandelbrot, *The Fractal Geometry of Nature.* New York: Freeman, 1982.

[15] N. C. Kenkel and D. J. Walker, "Fractals in the biological sciences," *Coenoses,* vol. 11, pp. 77-100, 1996.

[16] P. C. Ivanov, L. A. N. Amaral, A. L. Goldberger, S. Havlin, M. G. Rosenblum, Z. R. Struzik and H. E. Stanley, "Multifractality in human heartbeat dynamics," *Nature,* vol. 399, pp. 461-465, 1999.

[17] H. Yang, S. T. S. Bukkapatnam and R. Komanduri, "Nonlinear adaptive wavelet analysis of electrocardiogram signals," *Phys. Rev. E,* vol. 76, pp. 026214, 2007.

[18] J.-P. Leduc, "Spatio-temporal wavelet transforms for digital signal analysis," *Signal Processing,* vol. 60, pp. 23-41, 1997.

[19] J. F. Muzy, E. Bacry and A. Arneodo, "The multifracal formalism revisited with wavelets," *International Journal of Bifurcation and Chaos,* vol. 4, pp. 245-302, 1994.

[20] Y. Chen and H. Yang, "A comparative analysis of alternative approaches for exploiting nonlinear dynamics in heart rate time series," in *Proceedings of 2013 IEEE Engineering in Medicine and Biology Society Conference (EMBC),* Osaka, Japan, 2013.

[21] R. K. P. Zia, E. F. Redish and S. R. Mckay, "Making sense of the Legendre transform," *Am. J. Phys.,* vol. 77, pp. 614-22, 2009.

[22] J. F. Muzy, E. Bacry, A. Arneodo, "Wavelets and multifractal formalism for singular signals: Application to turbulence data," *Phys Rev Lett,* vol. 67, pp. 3515-3518, 1991.

[23] J. Eckmann, S. O. Kamphorst and D. Ruelle, "Recurrence Plots of Dynamical Systems," *Europhys. Lett.,* vol. 4, pp. 973, 1987.

[24] N. Marwan, R. M. Carmen, M. Thiel and J. Kurths, "Recurrence plots for the analysis of complex systems," *Phys. Rep.,* vol. 438, pp. 237-329, 2007.

[25] R. Sun and Y. Wang, "Predicting termination of atrial fibrillation based on the structure and quantification of the recurrence plot," *Med. Eng. Phys.,* vol. 30, pp. 1105-1111, 2008.

[26] J. P. Zbilut, N. Thomasson and C. L. Webber, "Recurrence quantification analysis as a tool for nonlinear exploration of nonstationary cardiac signals," *Med. Eng. Phys.,* vol. 24, pp. 53-60, 2002.

[27] N. Marwan and J. Kurths, "Nonlinear analysis of bivariate data with cross recurrence plots," *Phys. Lett. A,* vol. 302, pp. 299-307, 2002.

[28] N. Thomasson, T. J. Hoeppner, C. L. Webber Jr. and J. P. Zbilut, "Recurrence quantification in epileptic EEGs," *Phys. Lett. A,* vol. 279, pp. 94-101, 2001.

[29] Z. Wu, "Recurrence plot analysis of DNA sequences," *Phys. Lett. A,* vol. 332, pp. 250-255, 2004.

[30] F. Strozzi, J. Zaldivar and J. P. Zbilut, "Recurrence quantification analysis and state space divergence reconstruction for financial time series analysis," *Physica A,* vol. 376, pp. 487-499, 2007.

[31] H. Yang, S. T. S. Bukkapatnam and L. G. Barajas, "Local recurrence based performance prediction and prognostics in the nonlinear and nonstationary systems," *Pattern Recognit.,* vol. 44, pp. 1834-40, 2011.

[32] M. Siek and D. P. Solomatine, "Nonlinear chaotic model for predicting storm surges," *Nonlin. Processes Geophys.,* vol. 17, pp. 405-420, 2010.

[33] M. A. Riley and S. Clark, "Recurrence analysis of human postural sway during the sensory organization test," *Neurosci. Lett.,* vol. 342, pp. 45-48, 2003.

[34] W. C. Skamarock and J. B. Klemp, "A time-split nonhydrostatic atmospheric model for weather research and forecasting applications," *Journal of Computational Physics,* vol. 227, pp. 3465-3485, 2008.

[35] Q. Huang, "Physics-driven Bayesian hierarchical modeling of the nanowire growth process at each scale," *IIE Transactions,* vol. 43, pp. 1-11, 2011.

[36] N. E. Huang, Z. Shen, S. R. Long, M. C. Wu, H. H. Shih, Q. Zheng, N. Yen, C. C. Tung and H. H. Liu, "The empirical mode decomposition and the Hilbert spectrum for nonlinear and non-stationary time series analysis," *Proceedings of the Royal Society of London. Series A: Mathematical, Physical and Engineering Sciences,* vol. 454, pp. 903-95, 1998.

[37] Z. Wu and N. E. Huang, "A study of the characteristics of white noise using the empirical mode decomposition method," *Proceedings of the Royal Society of London. Series A: Mathematical, Physical and Engineering Sciences,* vol. 460, pp. 1597-1611, 2004.

[38] P. S. Addison, "Wavelet Transforms and the ECG: A Review," *Physiol. Meas.,* vol. 26, pp. 155-199, 2005.

[39] H. Yang, "Multiscale recurrence quantification analysis of spatial cardiac vectorcardiogram (VCG) signals," *IEEE Trans. Biomed. Eng.,* vol. 58, pp. 339-347, 2011.

[40] Y. Chen and H. Yang, "Self-organized neural network for the quality control of 12-lead ECG signals," *Physiological Measurement,* vol. 33, pp. 1399, 2012.

[41] H. Yang, C. Kan, G. Liu and Y. Chen, "Spatiotemporal differentiation of myocardial infarctions," *Automation Science and Engineering, IEEE Transactions on,* vol. 10, pp. 938-47, 2013.

[42] J. Malmivuo and R. Plonsey, *Bioelectromagnetism: Principles and Applications of Bioelectric and Biomagnetic Fields.* USA: Oxford University Press, 1995.

[43] D. Dubin, *Rapid Interpretation of EKG's: An Interactive Course.* Cover Publishing Company, 2000.

[44] G. D. Clifford, F. Azuaje and P. E. McSharry, *Advanced Methods and Tools for ECG Data Analysis.* London: Artech House, 2006.

[45] M. Stridh, L. Sormmol, C. Meurling and B. Olsson, "Frequency trends of atrial fibrillation using the surface ECG," in *Engineering in Medicine and Biology Society (EMBC), Proceedings of 1999 Annual International Conference of the IEEE,* Atlanta, GA, 1999.

[46] N. V. Thakor and Y. Zhu, "Applications of Adaptive Filtering to ECG Analysis: Noise Cancellation and Arrhythmia Detection," *Biomedical Engineering, IEEE Transactions on,* vol. 38, pp. 785-94, 1991.

[47] V. X. Afonso, W. J. Tompkins, T. Q. Nguyen and M. Shen Luo, "ECG Beat Detection Using Filter Banks," *Biomedical Engineering, IEEE Transactions on,* vol. 46, pp. 192-202, 1999.

[48] S. T. S. Bukkapatnam, R. Komanduri, H. Yang, P. Rao, W. C. Lih, M. Malshe, L. M. Raff, B. Benjamin and M. Rockley, "Classification of atrial fibrillation episodes from sparse electrocardiogram data," *Journal of Electrocardiology,* vol. 41, pp. 292-99, 2008.

[49] C. Li, C. Zheng and C. Tai, "Detection of ECG characteristic points using wavelet transforms," *Biomedical Engineering, IEEE Transactions on,* vol. 42, pp. 21-8, 1995.

[50] S. C. Saxena and V. Kumar, "QRS detection using new wavelets," *Journal of Medical Engineering and Technology,* vol. 26, pp. 7-15, 2002.

[51] C. Lin, Y. Du and Y. Chen, "Adaptive wavelet network for multiple cardiac arrhythmias recognition," *Expert. Syst. Appl.,* vol. 34, pp. 2601-11, 2008.

[52] F. M. Robert and R. J. Povinelli, "Identification of ECG Arrhythmias Using Phase Space Reconstruction," *Lecture Notes in Computer Science,* vol. 2168, pp. 411-23, 2001.

[53] A. L. Goldberger, L. Amaral, L. Glass, J. Haussdorff, P. C. Ivanov, R. Mark, J. Mietus, G. Moody, C.-K. Peng and H. E. Stanley, "PhysioBank, physiotoolkit, and physionet: Components of a new research resource for complex physiologic signals," *Circulation,* vol. 23, pp. e215-e220, 2000.

[54] K. Q. Shen, C. J. Ong, X. P. Li, Z. Hui and E. P. V. Wilder-Smith, "A Feature Selection Method for Multilevel Mental Fatigue EEG Classification," *Biomedical Engineering, IEEE Transactions on,* vol. 54, pp. 1231-1237, 2007.

[55] G. E. Dower, A. Yakush, S. B. Nazzal, R. V. Jutzy and C. E. Ruiz, "Deriving the 12-lead electrocardiogram from four (EASI) electrodes," *J. Electrocardiol.,* vol. 21, pp. S182-7, 1988.

[56] G. E. Dower and H. B. Machado, "XYZ data interpreted by a 12-lead computer program using the derived electrocardiogram," *J. Electrocardiol.,* vol. 12, pp. 249-61, 1979.

[57] D. Dawson, H. Yang, M. Malshe, S. T. S. Bukkapatnam, B. Benjamin and R. Komanduri, "Linear affine transformations between 3-lead (Frank XYZ leads) vectorcardiogram and 12-lead electrocardiogram signals," *J. Electrocardiol.,* vol. 42, pp. 622-30, 2009.

[58] H. Yang, M. Malshe, S. T. S. Bukkapatnam and R. Komanduri, "Recurrence quantification analysis and principal components in the detection of myocardial infarction from vectorcardiogram signals," in *Proceedings of the 3rd INFORMS Workshop on Data Mining and Health Informatics (DM-HI 2008),* Washington, DC, USA, 2008.